\documentclass[%
reprint,
twocolumn,
superscriptaddress,
showpacs,preprintnumbers,
nofootinbib,
aps,
prd
]{revtex4-1}

\usepackage{amsmath}
\usepackage{amssymb}
\usepackage{mathtools}
\usepackage{hyperref}
\usepackage{graphicx}
\usepackage{color}
\usepackage{comment}
\usepackage{subfigure}
\usepackage{dcolumn}

\usepackage[utf8]{inputenc}

\newcommand\MSbar{$\overline{\text{MS}}$ } 
\newcommand\dlT[1]{\eta_{#1}}

\begin{document}

\newcommand{\HIPetc}{\affiliation{
    Department of Physics and Helsinki Institute of Physics,
    PL 64, 
    FI-00014 University of Helsinki,
    Finland
}}

\newcommand{\UIUCAff}{\affiliation{
    Department of Physics, 
    University of Illinois, 
    Urbana, IL 61801, USA
}}

\newcommand{\umassAff}{\affiliation{
    Amherst Center for Fundamental Interactions,
    Department of Physics, \\
    University of Massachusetts,
    Amherst,
    MA 01003, USA
}}

\newcommand{\TDLIAff}{\affiliation{
Tsung-Dao Lee Institute and  School of Physics and Astronomy, Shanghai Jiao Tong University, 800 Dongchuan Road, Shanghai, 200240 China
}}

\newcommand{\CaltechAff}{\affiliation{Kellogg Radiation Laboratory, California Institute of Technology, Pasadena, CA 91125 USA
}}

\newcommand{\ITPAff}{\affiliation{
    Albert Einstein Center for Fundamental Physics,
    Institute for Theoretical Physics,
    University of Bern \\
    Sidlerstrasse 5,
    CH-3012 Bern,
    Switzerland
}}

\title{Nonperturbative analysis of the gravitational waves from a\\
first-order electroweak phase transition}

\author{Oliver~Gould}
\email{oliver.gould@helsinki.fi}
\HIPetc

\author{Jonathan~Kozaczuk}
\email{kozaczuk@illinois.edu}
\UIUCAff
\umassAff

\author{Lauri~Niemi}
\email{lauri.b.niemi@helsinki.fi}
\HIPetc

\author{Michael~J.~Ramsey-Musolf}
\email{mjrm@physics.umass.edu}
\umassAff
\TDLIAff
\CaltechAff

\author{Tuomas~V.~I.~Tenkanen}
\email{tenkanen@itp.unibe.ch}
\HIPetc
\ITPAff

\author{David~J.~Weir}
\email{david.weir@helsinki.fi}
\HIPetc

\date{\today}

\begin{abstract}

We present the first end-to-end nonperturbative analysis of the
gravitational wave power spectrum from a thermal first-order
electroweak phase transition (EWPT), using the framework of 
dimensionally reduced effective field theory and pre-existing
nonperturbative simulation results. We are able to show that a first-order EWPT
 in any beyond the Standard Model (BSM) scenario 
that can be described by a Standard Model-like effective theory at long distances
will produce gravitational wave signatures
too weak to be observed at existing and planned detectors. This implies that colliders are likely to provide the 
best chance of exploring the phase structure
of such theories, while transitions strong enough to be detected at gravitational wave experiments
require either previously neglected higher-dimension
operators or light BSM fields to be included in the dimensionally reduced effective theory
and therefore necessitate dedicated nonperturbative studies.
 As a concrete application, we analyze the real
singlet-extended Standard Model and identify regions of
parameter space with single-step first-order transitions, comparing
our findings to those obtained using a fully perturbative method. We discuss
the prospects for exploring the electroweak phase diagram in this model
at collider and gravitational wave experiments in light of our nonperturbative results.

\end{abstract}
\preprint{HIP-2019-5/TH}
\preprint{ACFI T19-04}
\maketitle

\section{Introduction}

The nature of electroweak symmetry breaking in the early Universe has important implications for cosmology and particle physics.
An electroweak phase transition (EWPT) could have given rise to
the observed matter/antimatter asymmetry of the
universe~\cite{Morrissey:2012db}, significantly affected the abundance of cosmic relics~\cite{Wainwright:2009mq}, or produced a stochastic
background of gravitational waves (GWs)~\cite{Caprini:2015zlo}. In the pure Standard Model (SM),
electroweak symmetry breaking is expected to have occurred at a crossover for 
the observed value of the Higgs mass
\cite{Kajantie:1995kf,Kajantie:1996mn,Kajantie:1996qd,Csikor:1998eu}.
However, many of its extensions predict much stronger
transitions, and in particular first-order phase transitions that may
have sourced gravitational
waves that can be observed at present day or future experiments~\cite{Caprini:2015zlo,Weir:2017wfa}.

In light of the successes of the LIGO and VIRGO
collaborations~\cite{Abbott:2016blz,TheLIGOScientific:2017qsa}, as
well as the LISA Pathfinder spacecraft~\cite{Armano:2016bkm}, a number
of space-based gravitational wave missions are now planned or in
preparation. Most notable of these is LISA~\cite{Audley:2017drz}, due to launch before 2034;
others, such as DECIGO~\cite{Kawamura:2011zz} and BBO~\cite{Harry:2006fi}, may follow. These space-based interferometers will have arm lengths
long enough to search for a stochastic GW background produced during the epoch of electroweak symmetry breaking in the early Universe, and therefore at a first-order electroweak phase transition.
There are also many future collider projects  
currently being discussed that have the potential of probing
the nature of the electroweak phase transition. In addition to the high-luminosity
LHC (HL-LHC)~\cite{Cepeda:2019klc}, these include proposals for $e^+e^-$ colliders such as
the ILC~\cite{Baer:2013cma}, FCC-ee~\cite{Benedikt:2018qee}, CEPC~\cite{CEPCStudyGroup:2018ghi} and CLIC~\cite{Linssen:2012hp}, as well
as higher-energy proton-proton colliders such as the High-Energy LHC (HE-LHC)~\cite{Zimmermann:2018wdi},
FCC-hh~\cite{Benedikt:2018csr} and SPPC~\cite{CEPC-SPPCStudyGroup:2015csa}. Various measurements at these machines
will provide valuable insight into the nature of electroweak symmetry breaking and complement the observations at gravitational wave experiments.

Given the rich experimental program aimed at exploring the electroweak phase transition, it is crucial to accurately characterize and make reliable predictions for the nature of the EWPT in extensions of the Standard Model. 
Due to the infamous Linde problem, perturbative methods in thermal
field theory are known to suffer from severe infrared divergences near
phase transitions \cite{Linde:1980ts,Gross:1980br}.  Furthermore, perturbative studies are often plagued by issues related to gauge dependence~\cite{Patel:2011th, Garny:2012cg}.
Establishing the existence of a strong first-order EWPT, and a precise determination of thermodynamic quantities related to it, therefore requires
nonperturbative methods. Such an approach is crucial for providing robust phenomenological predictions for gravitational wave and collider experiments.

A particular example of the limitations associated with perturbative calculations is the evaluation of the bubble nucleation rate, especially when the energy barrier between the vacua is radiatively induced. In this case thermal fluctuations are responsible for creating the barrier between the symmetric and broken phases (see Ref.~\cite{Chung:2012vg} for a discussion). For such transitions, the perturbative calculation of bubble nucleation requires a separation of scales between those fluctuations which are responsible for the energy barrier and those which describe the bubble nucleation~\cite{langer1974metastable,Berges:1996ib,Litim:1996nw}. One has to tread very carefully in order to avoid double counting fluctuations~\cite{Weinberg:1992ds,Gleiser:1993hf,Alford:1993br}, to self-consistently account for the spacetime dependence of the bubble~\cite{Surig:1997ne,Garbrecht:2015yza} and to maintain gauge invariance~\cite{Metaxas:1995ab,Baacke:1999sc,Patel:2011th,Garny:2012cg}.
Furthermore, explicit calculations taking into account the necessary separation of scales have shown a strong dependence on the precise matching scale, perhaps suggesting a breakdown of the usual semiclassical methods~\cite{Strumia:1999fv}. Beyond the leading order, calculating corrections to the bubble nucleation rate is a formidable task involving the
evaluation of functional determinants. In practice, these issues are rarely addressed by bubble  nucleation calculations in the literature. Instead practical shortcuts are adopted which, though not rigorously theoretically justified, are hoped to account for the relevant physics and provide the correct nucleation rate to within order of magnitude. For these reasons, it is important to know how well the usual approach performs quantitatively. This can only be answered by comparison to a fully self-consistent, nonperturbative approach \cite{Moore:2000jw}. In the paper at hand, we perform such a benchmark comparison.

In practice, the most convenient approach to bypassing the infrared
problems inherent in perturbative calculations  is provided by dimensional reduction
(DR)~\cite{Kajantie:1995dw,Appelquist:1974tg,Braaten:1995cm,Farakos:1994kx},
a technique for constructing an effective three-dimensional (3-d)
theory for the most infrared-sensitive light field modes. The
effective theory can then be studied on the lattice in an
infrared-safe
manner~\cite{Kajantie:1995kf,Kajantie:1996qd,Farakos:1995dn,Farakos:1994xh}. In the past, dimensional reduction has been applied to the
SM~\cite{Kajantie:1995kf,Kajantie:1996mn,Kajantie:1996qd}, the
MSSM~\cite{Laine:1996ms,Cline:1996cr,Farrar:1996cp,Cline:1997bm,Laine:1997qm,Laine:2000rm,Laine:2013raa}
the Two-Higgs-Doublet model
(2HDM)~\cite{Losada:1996ju,Andersen:1998br,Andersen:2017ika,Gorda:2018hvi}, 
as well as the real scalar singlet~\cite{Brauner:2016fla} (xSM) and triplet~\cite{Niemi:2018asa} ($\Sigma$SM) extensions of the SM (see also Ref.~\cite{Tenkanen:2018cwo}). 
Perturbative analyses of the EWPT in
these models -- usually based on the daisy-resummed one-loop effective
potential -- have been widely studied in e.g.~Refs.~\cite{Huber:2000mg,OConnell:2006rsp,Ahriche:2007jp,Profumo:2007wc,Espinosa:2011ax,
  Cline:2012hg, Damgaard:2013kva,Profumo:2014opa,Curtin:2014jma, Kozaczuk:2015owa, Huang:2015tdv, Damgaard:2015con, Curtin:2016urg, Beniwal:2017eik, Kurup:2017dzf, Alves:2018jsw, Alves:2018oct, Beniwal:2018hyi,Chen:2017qcz}
for the xSM, and for a selection of other models in e.g.~Refs.~\cite{Apreda:2001us, Dorsch:2014qja,Dorsch:2016nrg,Basler:2016obg,Basler:2017uxn,Patel:2012pi}.

Even in the effective 3-d description, a lattice study of the full
parameter space in BSM scenarios is computationally challenging. An
economic alternative to dynamical simulations becomes possible if the
new fields are sufficiently heavy, so that their effect on the
dynamics of the transition is suppressed. The BSM degrees of freedom
can then be integrated out in the process of dimensional reduction to obtain a
``SM-like effective'' 3-d theory whose phase structure is
well understood from the earlier lattice studies of
Refs.~\cite{Kajantie:1995kf,Kajantie:1996qd}; this strategy was first
adopted in Ref.~\cite{Cline:1996cr}. Although computationally inexpensive, this approach
comes with its own limitations. One cannot study
interesting multistep transitions \cite{Patel:2012pi, Blinov:2015sna} where new BSM
fields can be light near the transition, nor even one-step transitions
in which the new light BSM field plays an active, dynamical
role. Furthermore, integrating out the BSM fields can lead to reduced
accuracy when studying the dynamics of the transition, since the
effects of higher dimensional operators can be large but have been
neglected in all nonperturbative studies of the 3-d infrared effective theory
(cf. Ref.~\cite{Tenkanen:2018cwo}). Nevertheless, before performing
new simulations with additional dynamical BSM fields or higher dimension operators, it is
informative to investigate how much the existing nonperturbative
results can tell us about the theory. 

One objective of this study is to extract as much model-independent information from existing lattice results
as possible and apply them to gravitational wave predictions. The 3-d infrared SM effective field theory (EFT) is particularly simple
and we are able to determine the range of phase transition parameters relevant for computing the GW spectrum
accessible by this particular effective theory, in which effects from higher-dimension operators and light BSM fields 
are negligible. We find that a first-order EWPT in \textit{any} BSM theory that can be described by the minimal SM-like 3-d EFT upon integrating out the additional fields will lead to signatures that are \textit{undetectable} at any planned gravitational wave detector. That
is, one cannot get a sufficiently strong and slow transition from any theory which looks SM-like in the infrared. As a result, collider experiments
are likely to provide the most sensitive probe of the phase diagram in such scenarios. This also means that future lattice studies incorporating the effects of higher dimension operators or light BSM fields will be required in order to 
make theoretically sound predictions for gravitational wave experiments.

As a concrete application, the latter portion of this paper fleshes out these results in a specific BSM scenario, 
the real singlet-extended Standard Model (xSM)~\cite{Barger:2007im,Ashoorioon:2009nf,Robens:2015gla,Kanemura:2016lkz,Kanemura:2015fra,Beniwal:2017eik},
for which the required dimensional reduction has already been done in
Ref.~\cite{Brauner:2016fla}, but a systematic analysis of the
5-dimensional parameter space was not performed.  We
analyze the phenomenologically interesting parameter space
with a sufficiently heavy singlet, and make use of the
nonperturbative results of Ref.~\cite{Kajantie:1995kf} to deduce the
phase diagram of this model, comparing with the results of a traditional perturbative treatment as in Ref.~\cite{Chen:2017qcz}. We present the first nonperturbative results for
the gravitational wave power spectrum in the xSM with the level of rigor that can be achieved with existing
lattice results. As expected, the corresponding transitions are too weak and
cannot be realistically probed by gravitational wave experiments. Nevertheless, we use these results to perform a comparison between perturbative treatments and the nonperturbative predictions, finding reasonable agreement within systematic uncertainties 
for the various thermodynamic quantities relevant for the gravitational wave spectrum.
While a future nonperturbative study of this model will be required to 
make firm predictions for gravitational wave experiments, our analysis of the phase 
diagram provides robust targets for collider searches, which will have a realistic opportunity
to probe the first-order regions accessible by existing lattice results.

This article is organized as follows. In Section~\ref{sec:NP} we
review the setup for recycling pre-existing nonperturbative results
for determining the phase structure of a given BSM scenario. In
Section~\ref{sec:GW} we discuss in detail how to use the existing
nonperturbative results to predict gravitational wave signals from first-order
electroweak phase transitions in general BSM scenarios that map on to the infrared 3-d SM-like EFT.
In Section~\ref{sec:xSM} we focus on the real singlet extension
of the Standard Model and present our nonperturbative results for the phase diagram of the
model, comparing against predictions obtained from perturbation
theory. Sec.~\ref{sec:pheno} discusses collider and gravitational wave probes
of the phase diagram in this theory, applying the results of Sections~\ref{sec:NP}-\ref{sec:GW}. Finally, we conclude in Section
\ref{sec:conclusions}.

\section{Dimensional reduction and nonperturbative determination of
  the phase diagram}

\label{sec:NP}

The technique of dimensional reduction allows us to determine the phase
structure of a given model, and in particular establish the existence of first
order phase transition, in a theoretically sound manner.  The basic idea
of dimensional reduction is that at high temperatures, long-distance
physics decouples from the temporal direction, and the equilibrium
properties of the system can then be described by a spatial 3-d
theory. In the Matsubara formalism, this can be understood as a
consequence of a thermal scale hierarchy generated by the heat
bath. Namely, all non-zero Matsubara modes come with an effective mass
correction proportional to $\pi T$ -- a scale dubbed
\textit{superheavy} -- and their effect is thus suppressed compared to
modes with zero Matsubara frequencies. For more details regarding dimensional reduction,
see e.g.~Refs.~\cite{Kajantie:1995dw, Laine:1997qm, Laine:2016hma} and
\cite{Gorda:2018hvi,Niemi:2018asa}.

To illustrate the process of dimensional reduction in an extended scalar sector of the SM, we start with a four-dimensional (4-d) Euclidean Lagrangian in the imaginary-time formalism
\begin{align}
\label{eq:4-d-theory}
\mathcal L = \mathcal L (\phi, A_\mu, \psi, S, s),
\end{align}
where the 4-d fields depend on the imaginary time $\tau$ and spatial coordinates $\vec x$. The SM fields $\phi$, $A_\mu$ and $\psi$ -- representing the Higgs doublet, the gauge
fields and the fermions, respectively -- are accompanied by new scalars that we divide into two categories: by assumption the field $S$ is superheavy ($m_S \sim \pi T$) and the scalar $s$ is \textit{heavy}, meaning its mass is suppressed in a power counting sense ($m_s \sim g T$). We emphasize that here $m_S$ and $m_s$ are running masses in e.g.~the $\overline{\text{MS}}$ scheme, and not physical masses.
All fermionic modes, and bosonic hard modes, i.e.~those with
non-zero Matsubara frequency, have masses that are superheavy. 
Integrating out these superheavy field
modes (including the zero-mode of the scalar $S$) produces a bosonic 3-d theory with the Lagrangian
\begin{align}
\mathcal L_3 = \mathcal L_3 (\phi_3, A_i, A_0, s_3),
\end{align}
where the purely spatial 3-d fields (that correspond to zero modes of original fields) obtain masses that are
generally smaller than the superheavy scale.
The effects of the non-zero modes are captured by the parameters and
field normalizations of the effective theory, which are functions of
the temperature and physical quantities of the original 4-d
theory. Note that since the heat bath breaks four-dimensional
Lorentz invariance, the temporal components of the gauge fields $A_0$
are treated as Lorentz scalars in the effective 3-d theory.  Thermal
masses generated for these gauge field zero-modes are of order $g
T$, i.e.~at the heavy scale. In contrast, the spatial components of the non-Abelian gauge bosons remain massless. The scalar fields
undergoing a phase transition can also become very light, with masses formally of order $g^2 T$, due to a
cancellation between vacuum masses and thermal corrections. In the scenarios of interest below, it is assumed that only the Higgs doublet $\phi$ becomes light near the critical temperature. The EWPT is then driven by the Higgs field while the scalars $S,s$ are spectators. In order to affect the transition dynamics, the new scalars need to have sufficiently strong interactions with Higgs.  

\begin{figure}
\begin{center}
  \includegraphics[width=0.48\textwidth]{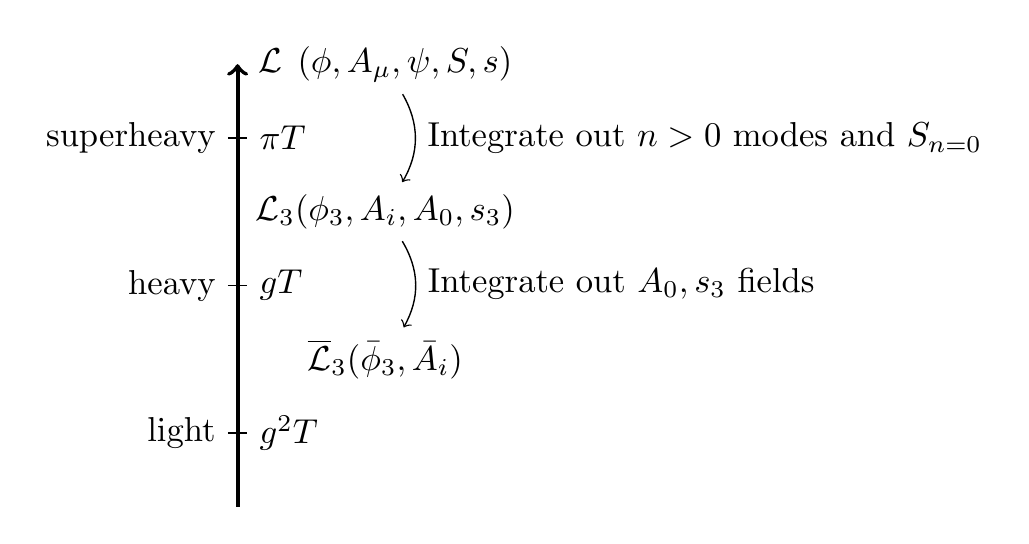}
\end{center}
\caption{A schematic illustration of the steps of dimensional
  reduction in the presence of \textsl{superheavy} and \textsl{heavy}
  BSM scalar fields $S$ and $s$, respectively. In the first step of
  DR, all fermionic modes, hard bosonic modes and all $S$ modes
   are integrated out, resulting in an effective 3-d theory of purely
  spatial zero-modes. In the second step of DR, the heavy scalars $A_0$
  and $s_3$ are integrated out, resulting in a simplified EFT of a
  doublet $\bar{\phi}_3$ and a gauge field $\bar{A}_i$ describing the long distance physics relevant for the phase transition.\label{fig:DR-fig}}
\end{figure}

In practice, fields belonging to the heavy scale can be integrated out as well, in the conventional
effective field theory sense. This results in a still simpler 3-d
theory
\begin{align}\label{eq:3dLagrangian}
\bar{\mathcal L}_3 = \bar{\mathcal L}_3 (\bar {\phi}_3, \bar{A}_i),
\end{align}
where the physics of the heavy scale is captured in the parameters and
field normalizations of the final light scale effective theory.  The
steps of dimensional reduction are illustrated in
Fig~\ref{fig:DR-fig}.  In the end, one arrives at an effective
theory solely for the infrared-sensitive zero modes.  Therefore the
procedure of constructing the effective theory, while perturbative in
nature, is completely infrared-safe and can be understood as a
sophisticated method for performing thermal resummation \cite{Farakos:1994kx,Farakos:1994xh} (see also Section 6 of Ref.~\cite{Laine:2016hma} for a more pedagogical discussion of this point). At the light
scale, the physics of the phase transition becomes inherently
nonperturbative due to infrared divergences in the effective
expansion parameter.  The theory can, however, be studied on the lattice.

The end result of dimensional
reduction for theories with only heavy and superheavy BSM physics is therefore 
a set of matching relations between the renormalized
parameters of the theory in Eq.~(\ref{eq:4-d-theory}) and a 3-d theory
containing the spatial $\mathrm{SU(2)}$ gauge field and a doublet
Higgs field,
\begin{equation}
  \label{eq:3dscalarpotential}
  \bar{\mathcal{L}}_{3,\text{scalar}} = D_i \bar{\phi}_3^\dag D_i \bar{\phi}_3 +
  \bar{\mu}_{h,3}^2 \bar{\phi}_3^\dag \bar{\phi}_3 + \bar{\lambda}_{h,3} (\bar{\phi}_3^\dag \bar{\phi}_3)^2,
\end{equation}
with gauge coupling $\bar{g}_3$. Note that nonperturbative effects of
the $\mathrm{U}(1)_Y$ gauge group were omitted in the original study
\cite{Kajantie:1995kf}, and we follow this same approach. Eq.~(\ref{eq:3dscalarpotential}) is the same 3-d theory to which the
Standard Model was mapped in Ref.~\cite{Kajantie:1995dw}, and then
studied nonperturbatively in
Refs.~\cite{Kajantie:1995kf,Kajantie:1996mn}. Furthermore, the
nucleation rate was studied nonperturbatively in
Ref.~\cite{Moore:2000jw}. 

Remarkably, the theory described by Eqs.~(\ref{eq:3dLagrangian})-(\ref{eq:3dscalarpotential}), which we subsequently refer to as 
the 3-d SM-like EFT, is universal in the
sense that it can be used to describe all extensions of the SM
as long as the new degrees of freedom remain heavy near the transition
after thermal mass corrections are taken into account. Furthermore,
the effects of the neglected higher dimensional operators,
such as $(\bar{\phi}_3^\dag \bar{\phi}_3)^3$, should remain small.
In many cases, BSM interactions with the Higgs
doublet generate operators of dimension six and higher (counting dimensions
as in 4-d) that can
have a significant impact on the transition dynamics
(cf. Ref.~\cite{Andersen:2017ika}), and so dropping these operators
results in reduced accuracy. In our results below, we will specify where
we expect these operators to become important.

In scenarios described by the 3-d SM-like EFT in the infrared, 
the dynamics are fully determined by only two
parameters,
\begin{equation}
x \equiv \frac{\bar{\lambda}_{h,3}}{\bar{g}_3^2} \quad \text{and} \quad
y \equiv \frac{\bar{\mu}_{h,3}^2}{\bar{g}_3^4}, \label{eq:xy}
\end{equation}
which are, as a result of dimensional reduction, functions of the
temperature and couplings of the full four dimensional theory.  The
tree-level value for the critical temperature $T_c$ is given by the
line $y = 0$. The true, fully nonperturbative critical line can be
obtained from a Monte Carlo simulation and turns out to be rather
close to the tree-level approximation \cite{Kajantie:1995kf}.
Furthermore, the condition $y \approx 0$ requires the thermal
correction to the doublet mass parameter to effectively cancel its
vacuum mass. This therefore fixes the critical temperature in our
analysis to be of the same order of magnitude as the physical Higgs
mass (this also approximately sets the scale at which
  gravitational waves will be emitted).  The strength of the transition
is then controlled by the parameter $x$, and the lattice study of
Ref.~\cite{Kajantie:1995kf} showed that for $0 < x < 0.11$ the
transition is of first-order, and becomes a crossover for $x > 0.11$
\cite{Rummukainen:1998as}.  Note that if, for a given choice of 4-d parameters, $x$ becomes negative, the tree-level potential of the effective
theory is unbounded from below and cannot be used for simulations. This generally
signals that the dynamical effects from the BSM fields are too large
to be neglected, or that the aforementioned higher-dimensional
operators should be included in the effective theory.

Summarizing, one can use existing nonperturbative results to study the electroweak 
phase diagram in BSM scenarios by performing the following steps:
\begin{enumerate}
\item Dimensionally reduce the full 4-d theory to the infrared 3-d theory.
\item In the parameter space of the theory that can be accurately described by the 3-d SM-like EFT (i.e.~where there are no additional light fields or important higher-dimension operator effects), compute $x$ and $y$. These quantities fully determine the 3-d dynamics.
\item Identify regions for which $0<x<0.11$ and $y\approx 0$. In this parameter space, nonperturbative simulations predict a first-order electroweak phase transition.
\end{enumerate}
 We stress again that the procedure above applies to transitions driven by the Higgs field and for which the heavy BSM fields are non-dynamical.
Extending our results to dynamical BSM fields requires retaining these fields in the effective 3-d theory; this, in turn will
require fresh nonperturbative simulations in the future.

The simple EFT described by Eqs.~(\ref{eq:3dLagrangian}) and (\ref{eq:3dscalarpotential}) can also be used to predict quantities relevant for the gravitational wave power spectrum arising from first-order transitions. In the following section, we describe how this can be done.

\section{The gravitational wave power spectrum}
\label{sec:GW}

At cosmological first-order phase transitions, excepting the case of very
large supercooling, gravitational waves are primarily produced by the
collisions of bubbles and the subsequent evolution of the resulting
fluid sound waves. Numerical simulations of the coupled field-fluid
system give the spectrum of gravitational waves as a function of
various equilibrium and dynamical properties of the transition
\cite{Hindmarsh:2017gnf}. The key quantities that we would like to determine
nonperturbatively are the strength of the
phase transition, $\alpha$, the phase transition temperature,
$T_*$, and the inverse phase transition duration, $\beta/H_*$
\cite{Caprini:2015zlo,Weir:2017wfa}. The latter two can be found once
one knows the bubble nucleation rate, $\Gamma$. In this section we
review how this can be done making use of
existing 3-d lattice results. There is one additional relevant
quantity: the bubble wall speed, $v_\mathrm{w}$, for which no
nonperturbative studies exist. In our study, we treat it as an input
parameter.

\subsection{Dimensional reduction for bubble nucleation} \label{sec:dr_bubbles}

Given a theory that has been dimensionally reduced as in Sec.~\ref{sec:NP} and predicting a first-order electroweak phase transition, we can 
also study bubble nucleation in the 3-d theory. The philosophy here is the usual one of dimensional reduction: 4-d
quantities are computed by matching to the analogous quantities
computed in the 3-d effective theory,
Eq.~\eqref{eq:3dscalarpotential}. Equilibrium properties of the 3-d
dimensionally reduced theory have been studied nonperturbatively in
Ref.~\cite{Kajantie:1995kf} and in Ref.~\cite{Moore:2000jw} the rate
of bubble nucleation was determined. This allows an end-to-end
nonperturbative description of the phase transition. The relevant
results are collected in Table \ref{tab:params}.

\begin{table*}

  \begin{ruledtabular}
    \begin{tabular}{dddc}
      \multicolumn{1}{c}{$x_c$} &
      \multicolumn{1}{c}{$y_c$} &
      \multicolumn{1}{c}{$\Delta \ell_3$} &
      Source \\
      \colrule
      0.01830 & 0.05904(56)  & 4.07(13)  &
      \cite{Kajantie:1995kf} \\
      0.036   & \dag0.030    & 1.265   &
      \cite{Moore:2000jw}     \\
      0.06444 & -0.00146(35) & 0.491(8) &
      \cite{Kajantie:1995kf} \\
      0.08970 & -0.01531(69) & 0.302(18) &
      \cite{Kajantie:1995kf} \\
      0.0983(15) & -0.018 & 0  &
      \cite{Rummukainen:1998as}
    \end{tabular}
  \end{ruledtabular}

  \caption{Nonperturbatively obtained 3-d phase transition parameters
    from Refs.~\cite{Kajantie:1995kf,Rummukainen:1998as,Moore:2000jw}
    (see also Ref.~\cite{Gurtler:1997hr}).  The entry marked with a
    dagger is interpolated based solely on values for other $x_c$.
    The final row shows the point at which the phase transition
    becomes second order, separating first-order transitions from
    crossovers. Note that the results in this table do not include the
    effect of the $\mathrm{U}(1)$ field, which makes the phase
    transition slightly stronger~\cite{Kajantie:1996qd}. This moves
    the critical line endpoint to $x_c \approx
    0.11$~\cite{Rummukainen:1998as}.  \label{tab:params}}

\end{table*}

Following Ref.~\cite{Weir:2017wfa} we define the phase transition
strength $\alpha$ as the ratio of the latent heat of the transition to
the radiation energy density in the symmetric phase at the time of transition
\begin{equation}
\alpha(T) \equiv \frac{30}{g(T) \pi^2} \frac{L(T)}{T^4},
\end{equation}
where $L(T)$ is the latent heat and $g(T)$ the number of relativistic
degrees of freedom. When BSM scalar eigenstates are sufficiently
heavy, they are not relativistic degrees of freedom at the EWPT and we
have $g(T \sim T_*) = 106.75$. Note that the
gravitational wave spectrum is determined by $\alpha(T_*)$, not
$\alpha(T_c)$. However, only $\alpha(T_c)$ has been determined
nonperturbatively, so we concentrate on this in what follows and approximate  $\alpha(T_*)\approx \alpha(T_c)$
which is typically justified absent significant supercooling.

The relation of the latent heat at $T_c$ to its 3-d counterpart,
$\Delta l_3$, can be found in Ref.~\cite{Kajantie:1995kf} and it reads
\begin{equation}
\frac{L(T_c)}{T^4_c} = \frac{g^6_3}{T_c^3} \left(\eta_y(T_c) - \eta_x(T_c) \frac{dy_c}{dx_c} \right) \Delta l_3, \label{eq:latent_heat_dr}
\end{equation}
where we have introduced the functions
\begin{equation}
 \dlT{x}(T)\equiv \frac{dx}{d\log T } \quad \text{and} \quad
 \dlT{y}(T)\equiv \frac{dy}{d\log T }. \label{eq:dr_beta_fns}
\end{equation}
For a given 4-d theory, these $\eta$-functions determine how the couplings
of the 3-d effective theory run with 
temperature. They are analogous to the usual $\beta$-functions of 
quantum field theory, in the sense that they describe the change
in parameters with respect to energy scale variations, though their appearance 
is connected to the process of DR.

For a generic BSM theory, thermal corrections to couplings arise first
at one-loop. For $y$, this corresponds to thermal corrections at 
$\mathcal{O(}g^{-2})$,
the thermal mass corrections, whereas for $x$ this leads to thermal corrections
first at $\mathcal{O(}g^2)$.
Thus the temperature dependence of $x$
is significantly weaker than that of $y$,
\begin{equation} 
 \dlT{x}(T) \ll \dlT{y}(T), \label{eq:eta_inequality}
\end{equation}
whereas $dy_c/dx_c=\mathcal{O(}1)$, hence
\begin{equation}
\frac{L(T_c)}{T^4_c} \approx \frac{g^6_3}{T_c^3} \dlT{y}(T_c)\Delta l_3.\label{eq:latent_heat_approx}
\end{equation}

In order to calculate the phase transition temperature and
inverse phase transition duration, one needs to know the rate of
bubble nucleation, $\Gamma$. The 4-d nucleation rate is related to its 3-d
counterpart by~\cite{Moore:2000jw},
\begin{equation}
\Gamma = \frac{T}{\sigma_{\mathrm{el}}} \left(\frac{g^2}{4\pi}\right)^5 T^4\ \Gamma_{3-d}(x,y),\label{eq:rate_dr}
\end{equation}
where $\sigma_{\mathrm{el}}\sim T/\log(1/g)$ is the non-Abelian
``color" charge conductivity, arising in the effective description of
the real-time dynamics of the long wavelength degrees of freedom
\cite{Bodeker:1998hm,Arnold:1998cy}. The powers of $g^2/(4\pi)$ in the
prefactor arise due to the relation between dimensionally reduced
lattice quantities and physical quantities. Finally,
$\Gamma_{3-d}(x,y)$ is the (dimensionless) rate of bubble nucleation
as calculated on the lattice in the 3-d effective theory. It depends
only on the 3-d parameters $x$ and $y$ (see Eq.~\eqref{eq:xy}).

The phase transition temperature, $T_*$, may be defined to be
the temperature at which only a fraction $1/\mathrm{e}\approx 0.37$ of
the universe remains in the symmetric phase. At this point the
following equality holds \cite{Enqvist:1991xw}
\begin{equation}
 \Gamma = \frac{1}{8\pi v_w^3}\left(H \frac{d\log \Gamma}{d\log T}\right)^4,\label{eq:nucleation}
\end{equation}
where $H$ is the Hubble parameter. Note that this condition is
different from the one-bubble-per-Hubble-volume condition, which
determines the very beginning of bubble
nucleation~\cite{Enqvist:1991xw,Anderson:1991zb}, yielding what
  is often termed the nucleation temperature $T_\mathrm{n}$. However,
  it is at the later time corresponding to $T_*$, when the universe
contains many bubbles, that bubble collisions occur, imprinting their
length-scale on the fluid sound waves and the resulting gravitational
wave spectrum. This distinction can lead to a factor of 2 difference
in $\beta/H_*$ and hence is important if one wants to make accurate
predictions.

The inverse duration of the phase transition is then determined by
\begin{align}
\frac{\beta}{H_*} & = -\frac{d\log \Gamma}{d\log T}\bigg|_{T=T_*} \nonumber \\
& \approx -\left(\dlT{x}(T_*)\frac{\partial}{\partial x} +
\dlT{y}(T_*)\frac{\partial }{\partial
  y}\right)\log\Gamma_{3-d}(x,y) \nonumber \\
&\approx -\dlT{y}(T_*)\frac{\partial }{\partial
  y}\log\Gamma_{3-d}(x,y), \label{eq:betaonH}
\end{align}
where all quantities are evaluated at the transition temperature\footnote{Note that in Ref. \cite{Moore:2000jw} the factor of $\eta_y$ was mistakenly dropped, leading to their result for $\beta/H_*$ being a factor of $\sim 4$ too small.}. On the last line we have used Eq.~\eqref{eq:eta_inequality}. From semiclassical calculations of the nucleation rate, one also expects the $x$ dependence of the rate to be weaker than the $y$ dependence, making the term we have kept even more dominant.

\begin{table}

\label{tab:sample-point}
  \begin{center}
      \begin{ruledtabular}
        \begin{tabular}{dddd}
          \multicolumn{1}{c}{$x_c$} &
          \multicolumn{1}{c}{$(y-y_c)$} &
          \multicolumn{1}{c}{$-\log\Gamma_{\mathrm{3-d}}$} &
          \mathrm{Source} \\
          \colrule
          0.036 & -0.00835 & 97.0\pm 0.7 & \text{\cite{Moore:2000jw}}\\
          0.036 & -0.00951 & 74.1\pm 0.6 &\text{\cite{Moore:2000jw}}\\
          0.036 & -0.01057 & 58.8\pm 0.7 &\text{\cite{Moore:2000jw}}\\
        \end{tabular}
      \end{ruledtabular}
  \end{center}

  \caption{The dimensionless rate of bubble nucleation at $x=0.036$ as
    calculated nonperturbatively in Ref.~\cite{Moore:2000jw}. In that
    reference $-\log\Gamma_{\mathrm{3-d}}$ is given in the last column
    of Table IV. In terms of their notation, $(y-y_c)\equiv \delta
    m^2/g^4T^2$. \label{tab:mr_rate}}

\end{table}

In Ref.~\cite{Moore:2000jw}, the quantity $\Gamma_{3-d}(x,y)$ was calculated for $x=0.036$ and is reproduced here in Table~\ref{tab:mr_rate}. It is given as a function of $(y-y_c)$, corresponding to $ \delta m^2/g^4T^2$ in the notation of Ref.~\cite{Moore:2000jw}. Fitting a straight line to their results, one finds
\begin{equation}
-\log\Gamma_{3-d}(0.036,y)\approx 240 + 1.72\times 10^4 (y-y_c),
\end{equation}
for $(y-y_c)\in [-0.01057,-0.00835]$, a range which is sufficient to
incorporate the value of $y$ at the transition temperature, $T_*$. Thus, for
this value of $x$,
\begin{align}
\frac{\beta}{H_*} &\approx 1.72\times 10^4\ \eta_y(T_*),\label{eq:beta_036} \\
\alpha(T_c) &\approx 2.2\times 10^{-3}\ \eta_y(T_c).\label{eq:alpha_036} 
\end{align}
Details of the specific 4-d theory only enter these expressions through $\eta_y$. Interestingly, by combining Eqs.~\eqref{eq:beta_036} and \eqref{eq:alpha_036}, one finds the relation
\begin{equation}
\frac{\beta}{H_*} \approx 7.8\times 10^{6} \alpha(T_c), \label{eq:alphabetaline} 
\end{equation}
 where we have used $\eta_y(T_*) \approx
 \eta_y(T_c)$. This defines a line in the
 $(\alpha(T_c),\beta/H_*)$ plane on which lie all 4-d theories which
 reduce to the SM-like 3-d EFT with $x=0.036$. A given 4-d theory's
 position along the line is determined solely by $\eta_y(T_c)$, 

An analogous relation will hold for other values of $x$: the ratio of
$\beta/H_*$ to $\alpha$ is strikingly independent of all short-range
4-d physics. Physically though, the dominance of 3-d, long-range
physics is no surprise. The tree-level 3-d theory is the result of
integrating out the superheavy and heavy modes of the 4-d
theory. However, the 3-d theory does not contain a tree-level
barrier. Hence one must integrate out some light degrees of freedom of
the 3-d theory to give the energy barrier between the vacua. The
degrees of freedom describing the bubble nucleation are then
necessarily lighter still. Thus, if any 4-d theory reduces to this 3-d
theory, the dynamics of its transition are largely independent of the
original 4-d theory, excepting the value of $x_c$ to which it maps.

\subsection{Implications for gravitational wave experiments}
\label{sec:gw_implications}

We are now in a position to make theoretically sound, nonperturbative
predictions for the gravitational wave spectrum resulting from a first-order electroweak phase transition in any BSM theory
that can be mapped into the EFT given by
Eqs.~(\ref{eq:3dLagrangian})-(\ref{eq:3dscalarpotential}). 

The general recipe for this is
described as follows, assuming Eq.~(\ref{eq:eta_inequality}) holds. 
First one solves the matching relations to find points in the 4-d
parameter space that give $x_c=0.036$, so that we can use the only
existing nonperturbative result for the bubble nucleation rate. Then by solving
Eq.~\eqref{eq:nucleation} one can find the phase transition temperature,
$T_*$, and finally evaluate Eqs.~\eqref{eq:beta_036} and
\eqref{eq:alpha_036} to give $\beta/H_*$ and $\alpha$. One can then
use the fit formula of Ref.~\cite{Hindmarsh:2017gnf} to determine the GW
power spectrum today. Variation of the bubble wall speed affects the
signal strength weakly, well within the range of other sources of
uncertainty, and so we simply set it to 1 in what follows.

\begin{figure*}
   \begin{center}
   \subfigure[]{\label{fig:GWs_a}
      \includegraphics[width=0.45\textwidth]{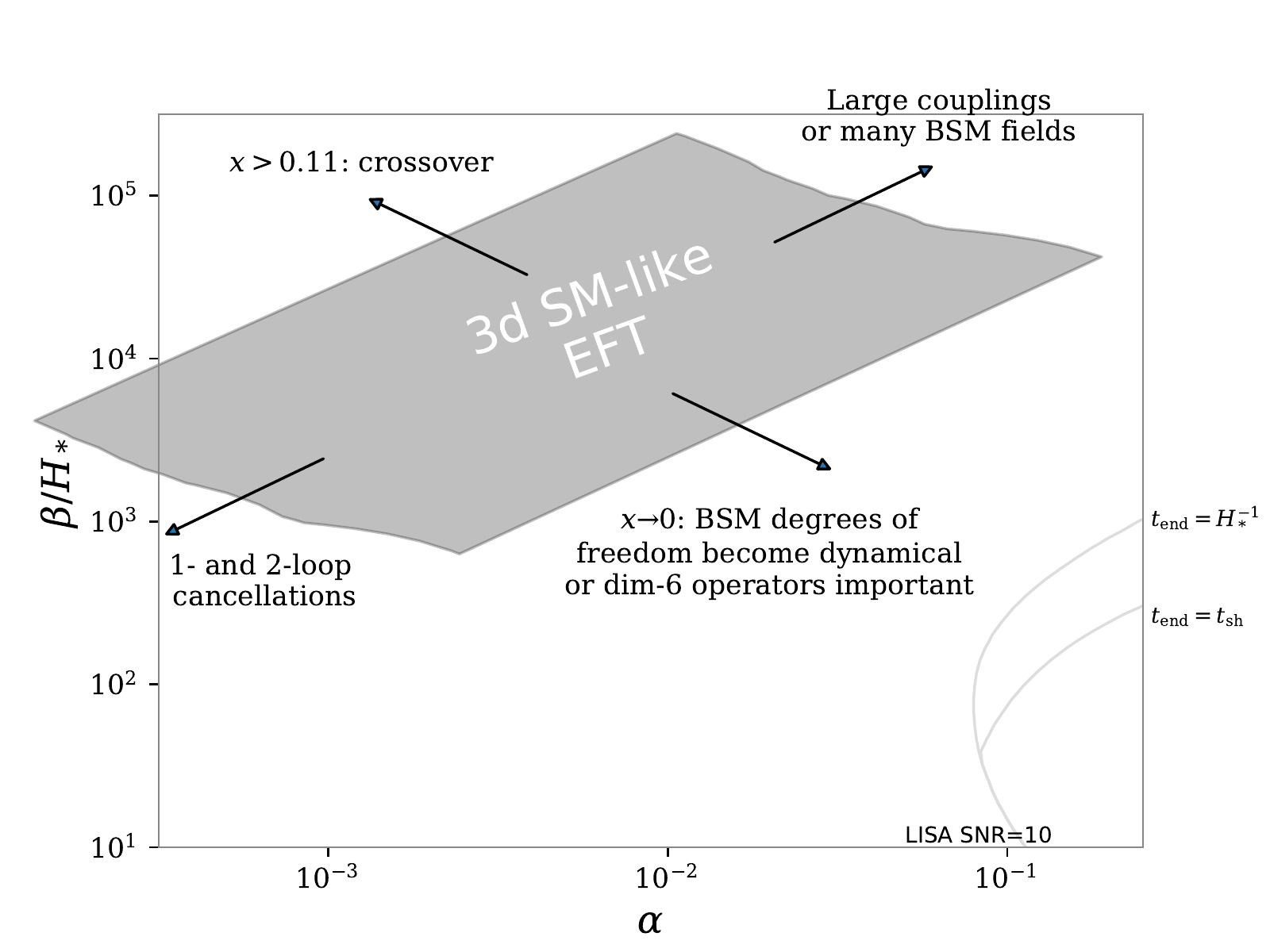}}
    \subfigure[]{\label{fig:GWs_b}
      \includegraphics[width=0.45\textwidth]{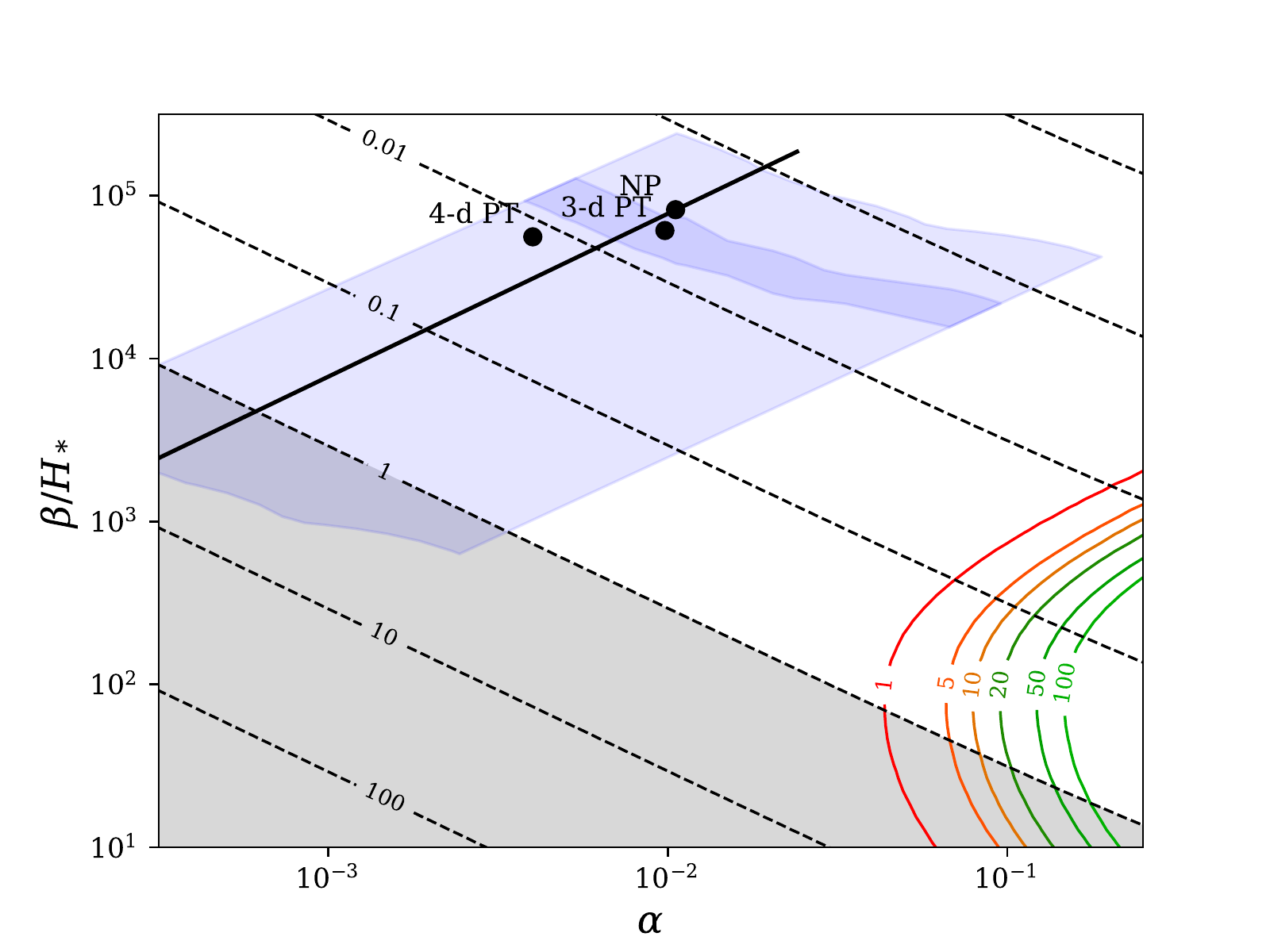}}
    
  \end{center}

  \caption{\label{fig:GWs}
    The entire region of gravitational wave parameter space onto
    which the 3-d Standard Model-like effective field theory can be
    mapped, along with prospects for detection. In panel (a), the corresponding region is shaded gray, along with explanations of the various limits to the
    effective field theory's validity. Also shown are LISA $\mathrm{SNR}=10$ sensitivity curves~\cite{Audley:2017drz} (assuming a five year mission duration) for two
    cases: one where the sound source is on until the shock formation time ($t_{\rm end}=t_{\rm sh}$),
    and the other where the sound source lasts the full Hubble time  ($t_{\rm end}=H^{-1}_*$). In panel (b), the parameter space mapping to the 3-d SM-like EFT is shown in light blue, while the black line shows the ranges of $\alpha$ and $\beta/H_*$ that can be determined  nonperturbatively 
    from the results of Ref.~\cite{Moore:2000jw}. Additional LISA SNR curves are shown in (b), assuming that the sound source lasts for a Hubble time (the
    dashed contours give the shock formation time, and only in the
    shaded region is the shock formation time longer than a Hubble
    time). Note that other gravitational wave experiments are not expected to provide additional sensitivity to the SM-like 3-d EFT regions.
     Finally, the dark blue region in (b) is the area of relevance for the real
    singlet model (xSM), as discussed in Sec.~\ref{sec:pheno}. The points shown correspond to the benchmark comparison
    of our nonperturbative analysis and the predictions of 3-d and 4-d perturbation theory. The sensitivity curves assume $v_w =1$ and $T_*=140$ GeV. The latter is the appropriate choice for transitions described by the 3-d SM-like theory, as the magnitude of $T_*$ is primarily set by the Higgs mass absent other light degrees of freedom or large higher-dimension operator effects.
    This figure shows that the 3-d SM-like EFT and existing lattice results cannot be reliably used
    to study electroweak phase transitions predicting an observable gravitational wave signal.}
    \end{figure*}

The aforementioned analysis, together with generic field theoretic arguments, yields a well-defined range of the gravitational wave parameters,
$\alpha$, $\beta/H_*$, and $T_*$ accessible by theories matching on to 
the 3-d SM-like EFT.
We present these results in Fig.~\ref{fig:GWs}. The gray shaded region in Fig.~\ref{fig:GWs_a} and the light blue region Fig.~\ref{fig:GWs_b}  can be reliably described by the infrared 3-d SM-like EFT. In Fig.~\ref{fig:GWs_a} we specifically state why this EFT description fails in different regions, as detailed below. The black line Fig.~\ref{fig:GWs_b}
corresponds to $\alpha$ and $\beta/H_*$ values accessible by existing nonperturbative calculations. For reference, we also show the expected sensitivity of LISA~\cite{Audley:2017drz} to this parameter space.
In Fig.~\ref{fig:GWs_a}, the region predicting a signal-to-noise ratio (SNR) of 10 at LISA is indicated in the bottom right corner. Sensitivity curves assuming a five year mission duration are shown for two
cases: one where the sound source is on for the shock formation time, and the other where the sound source lasts the full Hubble time (see Ref.~\cite{Ellis:2018mja} for a more detailed discussion of these distinct cases). In Fig.~\ref{fig:GWs_b}, we show additional LISA SNR curves assuming that the sound source lasts for a Hubble time. The dashed contours give the shock formation time, and only in the shaded region is the shock formation time longer than a Hubble time. In both panels we set $T_*=140$ GeV and $v_w=1$. Neither the sensitivity curves nor the region for which the SM-like EFT are valid are particularly sensitive to $T_*$; the scale is implicitly set by the physical Higgs mass.

The regions accessible by the 3-d SM-like EFT and existing nonperturbative results in Fig.~\ref{fig:GWs} are determined as follows. First let us consider the range of allowed  $\eta_y$. 
This quantity, together with $x_c$, determines $\alpha(T_c)$ through Eq.~(\ref{eq:latent_heat_dr}) and $\beta/H_*$ through Eq.~(\ref{eq:betaonH}) (for $\eta(T_*)\approx \eta(T_c)$).
The value of $\eta_y(T_c)$ is generically of order
$\mathcal{O(}g^{-2})$, absent new physics with large couplings. Its SM value for a $125~$GeV Higgs, $\eta_y^{\mathrm{SM}}(T_c)\approx 4.4$,
receives corrections from BSM physics,
\begin{equation}
 \eta_y(T_c) = \eta_y^{\mathrm{SM}}(T_c) + \Delta \eta_y^{\mathrm{BSM}}(T_c),
\end{equation}
where $\Delta \eta_y^{\mathrm{BSM}}(T_c)$ contains corrections which 
go to zero in the SM limit. In principle $\Delta \eta_y^{\mathrm{BSM}}(T_c)$ 
can be positive or negative, though all leading-order terms in 
$\eta_y^{\mathrm{SM}}(T_c)$ are positive definite.

There is a well-defined range of $\eta_y(T_c)$ expected in generic BSM theories
mapping on to the 3-d SM-like EFT. First, note that If $\Delta \eta_y^{\mathrm{BSM}}(T_c)$ is negative, $\eta_y(T_c)$ is reduced
with respect to its SM value, leading to weaker transitions
with a longer phase transition duration. 
In special cases such that the SM and 
BSM contributions cancel, one can in principle get very weak transitions. 
However, to achieve $\eta_y(T_c)\ll\eta_y^{\mathrm{SM}}(T_c)$ requires
order-by-order cancellations between SM and BSM physics and hence is 
fine-tuned. Any generic BSM physics which 
does not lead to such fine-tuned cancellations
at more than the leading order in the loop expansion will satisfy
$\eta_y(T_c) \gtrsim (g^2/4\pi)\eta^{\mathrm{SM}}_y(T_c)\approx 0.15$.
Here we have conservatively taken the loop expansion parameter to be $\sim g^2/4\pi$:
the presence of any larger couplings would strengthen this approximate lower bound.
Furthermore, such small $\eta_y(T_c)$, shown as the bottom left arrow in Fig.~\ref{fig:GWs_a},
lead to unobservably, weak transitions from 
the perspective of their gravitational wave signal. 

Conversely, 
if $\Delta \eta_y^{\mathrm{BSM}}(T_c)$ is positive, one finds stronger 
transitions with shorter phase transition durations. The size of 
$\Delta \eta_y^{\mathrm{BSM}}(T_c)$ is nevertheless constrained by 
perturbativity. The perturbative loop expansion breaks down when couplings 
become large, limiting $\Delta \eta_y^{\mathrm{BSM}}(T_c)$ to be formally $\mathcal{O}(1/g^4)$ or smaller.
This implies $\eta_y(T_c) \lesssim 10$ as a reasonable upper bound from perturbativity.
One possible caveat would be to instead add a large number
of weakly coupled degrees 
of freedom to increase $\Delta \eta_y^{\mathrm{BSM}}(T_c)$.
Though it is 
more subtle in this case, a sufficiently large number of degrees 
of freedom will also lead to a breakdown of perturbativity.
In either case, if $\Delta \eta_y^{\mathrm{BSM}}(T_c)$ is large,
the phase transition
duration becomes very short, shown as the top right arrow in 
Fig.~\ref{fig:GWs_a}. This pushes the peak of the
gravitational wave spectrum to higher frequencies, which 
are unfavorable for gravitational wave experiments.

From this reasoning, we conclude that any generic BSM physics which couples perturbatively 
to the Higgs and does not lead to nontrivial cancellations at more than one
loop order will satisfy $\eta_y(T_c)\in [0.15,10]$. Assuming this
we find that, for $x=0.036$,
Eqs. \eqref{eq:beta_036} and \eqref{eq:alpha_036} trace out the thick
black line shown in Fig.~\ref{fig:GWs_b}.

At other values of $x$, we do not have a nonperturbative determination
of the phase transition parameters. In this case, we use a fully
perturbative approach utilizing the DR (3-d PT). In the 3-d
PT approach, our calculations are performed at two-loop order and in
the Landau gauge, cf. Ref.~\cite{Gorda:2018hvi}. In the calculation of
$\Gamma_{3-d}$, we follow the perturbative semiclassical analysis
outlined in Ref. \cite{Moore:2000jw}, including the effects of wave
function renormalization. This gives the tunneling action accurate to
next-to-leading order in $x$. Note that the semiclassical prefactor
contributes at next-to-next-to-leading order in
$x$~\cite{Weinberg:1992ds,Moore:2000jw}.

Larger values of $x$ result in weaker first-order transitions and 
shorter phase transition durations, until
$x=0.11$ where the transition is a crossover and hence there is no
bubble nucleation. This corresponds to the upper left arrow in Fig.~\ref{fig:GWs_a}.
 Conversely, for smaller values of $x$, the
phase transition becomes stronger and longer in duration, hence
smaller $x$ is the interesting region for gravitational wave
production (bottom right arrow in Fig.~\ref{fig:GWs_a}). 
However, there is a limit to how small one can reliably go
in $x$, as the dimension-6 operators that have been dropped in the DR
can become comparable to, or larger than, the dimension-4 term, which
is proportional to $x$. The dominant higher dimensional operator is 
$c_6(\phi_3^\dagger \phi_3)^3$. The leading SM contribution to $c_6$
is $\mathcal{O(}y_t^6/(4\pi)^3)$, where $y_t$ is the top Yukawa coupling
(see Appendix \ref{sec:DR-accuracy}). There will also be contributions
from any BSM physics that couples to the
Higgs. For the effect of the dimension-6 operator to be small, the
following strong inequality should hold,
\begin{equation}
 \bar{\lambda}_{h,3}\langle (\bar{\phi}_3^\dagger\bar{\phi}_3)^2\rangle_c \gg 
 c_{6}\langle (\bar{\phi}_3^\dagger\bar{\phi}_3)^3\rangle_c,
\end{equation}
where the expectation values refer to the volume averaged 
condensates, evaluated at the critical point (see, for example, Ref. \cite{Farakos:1994xh}).
Calculating these 
condensates using the two-loop effective potential,
which should be reasonably accurate at small $x$, and taking 
$c_6$ to be its Standard Model value, we arrive at $x \gg 0.01$.
However, it should be remembered that the precise condition will
depend on BSM contributions to $c_6$. In practice, such small values of $x$ are 
likely overoptimistic and encourage us to investigate an extended
3-d EFT including higher dimensional operators in the future.

Using the 3-d PT approach to calculate $\Delta l_3$ and $\Gamma_{3-d}$
allows us to trace out the entire region on the $(\alpha,\beta/H_*)$ plane that
can be described by the SM-like 3-d EFT in Eqs.~\eqref{eq:3dLagrangian}-
\eqref{eq:3dscalarpotential}, as argued above. Considering {$x\in[0.01,0.06]$ and the aforementioned range of
$\eta_y(T_c)$, we arrive at the gray and light blue regions of Figs.~\ref{fig:GWs_a} and  \ref{fig:GWs_b}, repsectively.
Here we have chosen the upper value $x=0.06$ since larger values 
are not accurately described by our 3-d PT approach and our next-to-leading order
calculation of the bubble nucleation rate becomes numerically challenging in this regime.
Nevertheless values in the range $0.06<x<0.11$ 
give only very weak first order phase transitions.
Any BSM theory which reduces to the 3-d
SM-like EFT and gives a reasonably strong first-order phase transition
will lie in the light blue region of Fig.~\ref{fig:GWs_b}. The dark blue region in Fig.~\ref{fig:GWs_b} is for the specific case
of the singlet extended Standard Model (xSM), which we discuss further in Section \ref{sec:ssb_gws}.

From Fig.~\ref{fig:GWs} we can reasonably conclude that, in
the region where the effects of heavy BSM fields on the EWPT can be
captured by the 3-d SM-like EFT, a given model does not predict a sufficiently strong
transition to produce an observable gravitational wave signal at LISA. In fact, 
we have verified that \emph{no planned gravitational wave experiment is expected 
to be sensitive to the gray and light blue regions of Figs.~\ref{fig:GWs_a} and~\ref{fig:GWs_b}}, respectively.
In order to explore the phase diagrams of theories mapping onto these regions,
other experimental approaches, such as searches for BSM physics at colliders,
will be required. We will see how this plays out in the specific case of the real singlet-extended SM below. For
stronger transitions as required for an observable GW signal there are two options: either the BSM field must
play an active, dynamical role or higher dimension operators in the 3-d
EFT must be included. A nonperturbative description will therefore
require new 3-d lattice simulations including the dynamical BSM field,
or inclusion of higher-dimension operators involving the scalar doublet
(see Fig.~\ref{fig:GWs}).

\section{Application to the Standard Model with real singlet scalar}
\label{sec:xSM}
Up until this point our analysis has been quite general. As an illustration of 
our methods, we apply our nonperturbative analysis to the specific case of the often-studied
real singlet extension of the Standard Model (xSM)
~\cite{Barger:2007im, Profumo:2007wc, Ashoorioon:2009nf, Espinosa:2011ax, Robens:2015gla,Kanemura:2016lkz,Kanemura:2015fra,Chen:2017qcz, Beniwal:2017eik}.  
Using the dimensional reduction of
Ref.~\cite{Brauner:2016fla}, in this section we first investigate the 5-dimensional
parameter space of the model to determine the phase structure in the
regions where the singlet is sufficiently heavy to be integrated out. We will then
compute the gravitational wave spectrum nonperturbatively in Sec.~\ref{sec:pheno}, comparing
with perturbative results, and discuss the prospects for exploring the phase diagram of this scenario
at collider experiments.

\subsection{Parameterization of the model}
\label{sec:parametrisation}

The most general scalar Lagrangian involving an additional real singlet scalar field can be written as~\cite{Barger:2007im, Profumo:2007wc, Espinosa:2011ax, Chen:2014ask, Chen:2017qcz}
\begin{multline}
\label{eq:scalarpotential}
  \mathcal{L}_\text{scalar} = (D_\mu \phi)^\dag (D_\mu \phi) - \mu^2
  \phi^\dag \phi + \lambda (\phi^\dag \phi)^2 \\
  + \frac{1}{2}
  (\partial_\mu \sigma)^2 + \frac{1}{2} \mu_\sigma^2 \sigma^2 + b_1
  \sigma + \frac{1}{3} b_3 \sigma^3 + \frac{1}{4}b_4
  \sigma^4 \\
  + \frac{1}{2} a_1 \sigma \phi^\dag \phi + \frac{1}{2}
  a_2 \sigma^2 \phi^\dag \phi,
\end{multline}
where $D_\mu$ is the usual covariant derivative for the Higgs doublet,
and parameters are renormalized in the \MSbar scheme. In order to
relate these \MSbar parameters to physical quantities, we
reparametrize the potential as follows. After electroweak symmetry
breaking, we assume that the singlet field $\sigma$ has a
zero-temperature vev $\langle \sigma\rangle =0$, and that the Higgs
vev is $\langle \phi^\dag \phi \rangle = v^2/2 $ . This leads to the
following relations in the electroweak vacuum:
\begin{align}
\mu^2 = v^2 \lambda,  \quad b_1 = -a_1 \frac{v^2}{4}.
\end{align}
Furthermore, by diagonalizing the potential and solving for the mass
eigenvalues -- with the eigenstates denoted $h_1$ and $h_2$ -- the
parameters $\lambda, \mu$ and $a_1$ can be eliminated in favor of
physical masses $m_{1}, m_{2}$ and the mixing angle $\theta$.
We assume that $h_1$ is the measured Higgs boson with mass $m_1 = 125$
GeV. Finally, instead of the quartic portal coupling $a_2$, we take the trilinear $(h2, h2, h1)$ coupling $\lambda_{221}$ as an input parameter, 
which is one of the couplings relevant for scalar pair production processes at colliders~\cite{Curtin:2014jma, Craig:2014lda, Chen:2014ask, Chen:2017qcz}.
 We treat the singlet self-couplings
$b_3$ and $b_4$, given at a fixed \MSbar scale, as input parameters. 
As a subset of the general model parameter space, we will also consider the 
case in which $\sigma\rightarrow - \sigma$ under a discrete symmetry so that
$a_1$, $b_1$, $b_3$ and $\sin \theta$ are all set to zero in a technically natural way. 
We refer to this as the $Z_2$-symmetric xSM.

In scalar extensions of the SM, first-order phase transitions are
often associated with large portal couplings to the Higgs doublet.
We therefore expect zero-temperature effects from vacuum
renormalization to have a significant effect on the properties of the
EWPT. To account for these corrections, we employ a one-loop
renormalization procedure in the $T=0$ vacuum similar to that of
Refs.~\cite{Kajantie:1995dw,Laine:2017hdk,Niemi:2018asa}. In short,
the running parameters are related to observables at an initial scale,
which we take to be an average of the scalar masses, by requiring that
the physical masses match poles of the loop-corrected
propagators. Technical details of this calculation are presented in
Appendix~\ref{sec:vacuum_renormalization}.

\subsection{Identifying regions with a first-order transition}
\label{sec:identifying_fopt}

\begin{figure}
\begin{center}
  \includegraphics[width=0.5\textwidth]{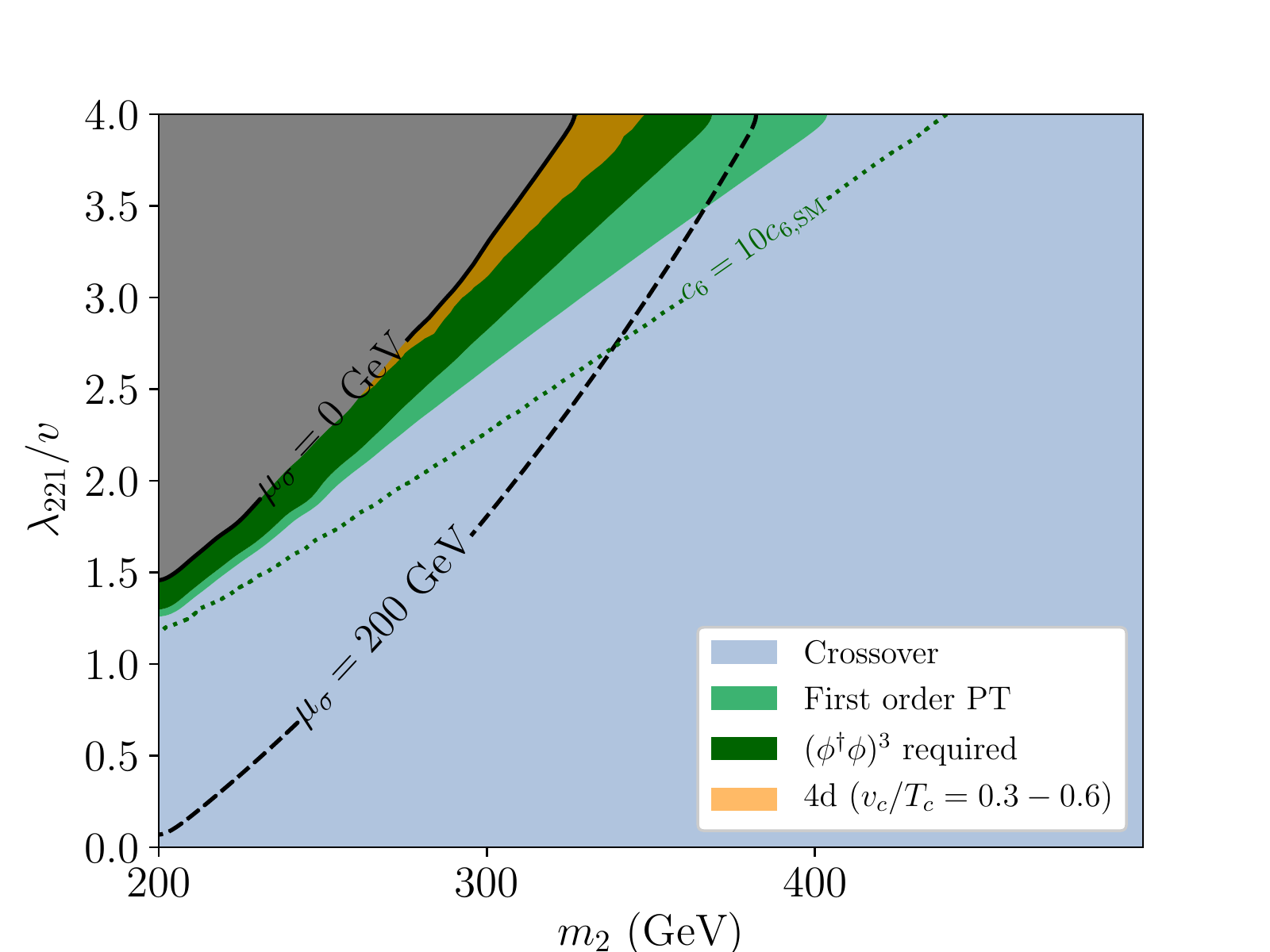}
\end{center}

\caption{\label{fig:Z2} Phase structure of the xSM in the $Z_2$ symmetric limit. Our nonperturbative approach predicts a first-order electroweak phase transition in the light green region. The darker green shaded region features $x<0$ so that the higher dimension operator $(\phi^\dagger \phi)^3$ must be kept in the dimensionally reduced theory to resolve one-step transitions. Furthermore, above the dotted green contour, the dimension-6 operator coefficient, $c_6$, becomes larger than 10 times the corresponding SM value, $c_{6, {\rm SM}}$, and the neglected  $(\phi^\dagger \phi)^3$ term can already have significant effects. The gray shaded region corresponds to where the singlet mass parameter $\mu^2_{\sigma}<0$ so that the superheavy dimensional reduction completely breaks down. The parameter space in which 4-d perturbation theory predicts a first-order transition with $v_c/T_c = 0.3-0.6$ is shaded orange. We also show contours indicating the size of the singlet \MSbar mass parameter $\mu_\sigma$. When $\mu_\sigma \lessapprox g T \approx 100$ GeV, the superheavy singlet approximation becomes compromised. A first-order EWPT is robustly excluded for small values of $\lambda_{221}$, where our nonperturbative treatment is well-justified.}
\end{figure}

We are interested in parameter space across which the singlet can be reasonably approximated as superheavy so that the dimensional reduction of Ref.~\cite{Brauner:2016fla} and the results of Sections~\ref{sec:NP}-\ref{sec:GW} can be applied. The singlet field in this case is non-dynamical at the transition and integrated out. Since this involves effectively replacing the singlet background field by a solution to its equation of motion (EOM), transitions between different singlet vacua will not be captured by this approach, and so in the superheavy regime one is restricted to considering one-step transitions across which the singlet VEV does not change significantly. This approach is therefore expected to capture the ``weakest'' transitions (i.e.~those with the smallest latent heat or order parameter). As singlet-induced higher dimension operators are required to be negligible in order to match on to the 3-d SM-like EFT in the infrared, the dominant effect strengthening the electroweak phase transition will be the reduction of the effective Higgs quartic coupling, with the barrier separating the phases generated radiatively in the EFT. This corresponds to the same mechanism responsible for driving the EWPT first-order in the Standard Model by reducing the Higgs mass. The reduction of the effective quartic coupling is primarily due to tree- (loop-) level effects in the non-$Z_2$ ($Z_2$) cases, respectively~\cite{Profumo:2007wc}. This can be seen explicitly in the expression in Ref.~\cite{Brauner:2016fla} for the 3-d Higgs quartic coupling after integrating out the superheavy degrees of freedom, $\lambda_{h,3}$, in terms of the model parameters:
\begin{equation}\label{eq:lam3}
\frac{\lambda_{h,3}}{T}=\lambda - \frac{a_1^2}{8\mu^2_{\sigma}} - \frac{a_1^2 b_3 b_1}{4\mu^6_{\sigma}} + \frac{a_1 a_2 b_1}{2\mu^4_{\sigma}} + {\rm loop \, effects}.
\end{equation}
The presence of the $Z_2$-breaking couplings $a_1, b_1, b_3$ lead to a tree-level reduction of $\lambda_{h,3}$, whereas absent these terms the leading effect starts at $\mathcal{O}(g^4)$ in the power counting scheme used in the dimensional reduction of Ref.~\cite{Brauner:2016fla}.

We would like to use our nonperturbative framework to ascertain where in singlet model parameter space the aforementioned effects are large enough to drive the EWPT first-order. Let us first consider the $Z_2$-symmetric limit of the model. In this case, $\sin \theta =0$, $b_3=0$, and we can simply vary $m_2$ and $\lambda_{221}$. We fix $b_4=0.25$, as the properties of the transition are not very sensitive to singlet self-interactions. Our nonperturbative results are shown in Fig.~\ref{fig:Z2}. A first-order electroweak phase transition is predicted in the light green region. In the darker green region, the 3-d parameter $x<0$, so that higher-dimensional operators are required to probe one-step transitions nonperturbatively. We also show where the neglected singlet-induced dimension-6 operator coefficient, $c_6$, becomes larger than 10 times the corresponding SM value, $c_{6, {\rm SM}}$, assuming $T\approx 140$ GeV. An approximate expression for $c_6$ can be found in Eq.~(\ref{eq:c6}). As discussed in Appendix~\ref{sec:DR-accuracy}, for $c_6 \gtrsim  10 c_{6,{\rm SM}}$ the neglected higher dimension operators are expected to have $\mathcal{O}(10\%)$ or larger effects on the 3-d Higgs VEV in the first-order transition regions (in 3-d perturbation theory using Landau gauge), and so our nonperturbative analysis already becomes less reliable for $\lambda_{221}$ above this contour. In the gray shaded region, the singlet mass parameter $\mu^2_{\sigma}<0$ and the superheavy dimensional reduction completely breaks down. We also show a contour for which the singlet mass parameter $\mu_{\sigma}=200$ GeV, above which our assumption $\mu_{\sigma}\sim \pi T$ is already questionable.

 As can be seen from Fig.~\ref{fig:Z2}, most of the first-order phase transition parameter space accessible to current nonperturbative studies lies in the region where the DR and neglect of higher dimension operators is not well-justified. For larger masses the superheavy approximation improves, however the portal couplings required for a first-order transition becomes larger for heavier singlets and therefore higher-loop zero-$T$ effects become non-negligible.  Below the $\mu_\sigma=200$ GeV and $c_6 \gtrsim  10 c_{6,{\rm SM}}$ contours, the superheavy singlet DR and neglect of higher-dimension operators is justified, and a first-order electroweak phase transition is robustly excluded by our nonperturbative results for small $\lambda_{221}$. This validates the conclusions drawn from perturbation theory in e.g.~Refs.~\cite{Craig:2014lda, Curtin:2014jma}. Note also that similar results for the phase diagram in the real triplet-extended SM were obtained in Ref.~\cite{Niemi:2018asa}.

\begin{figure*}[!t]

  \begin{center}

    \subfigure[\; $m_2 = 240$ GeV, $b_4 = 0.25$, $b_3 = 0$]{
      \includegraphics[width=0.45\textwidth]{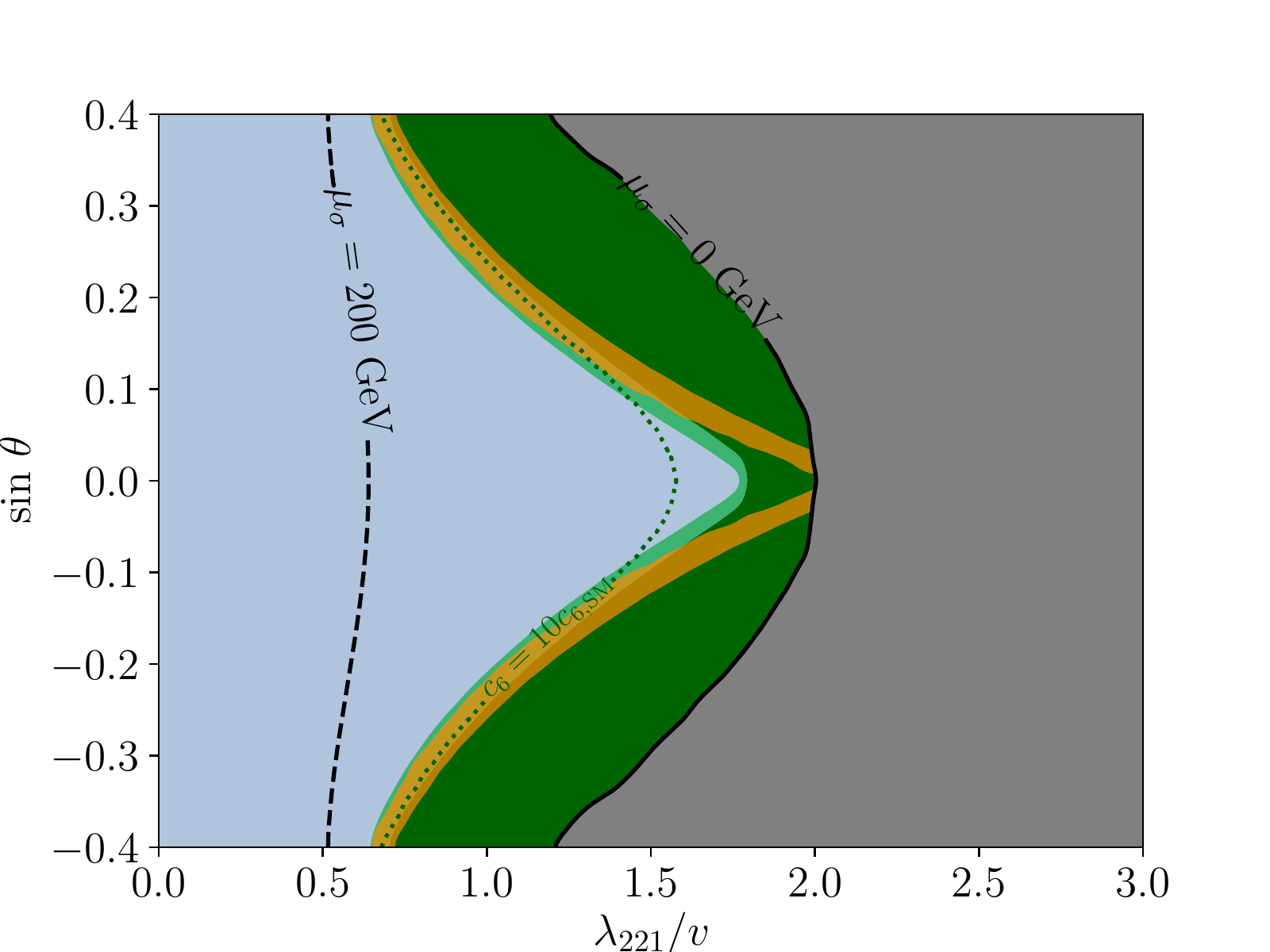}}
    \subfigure[\; $m_2 = 400$ GeV, $b_4 = 0.25$, $b_3 = 0$]{
      \includegraphics[width=0.45\textwidth]{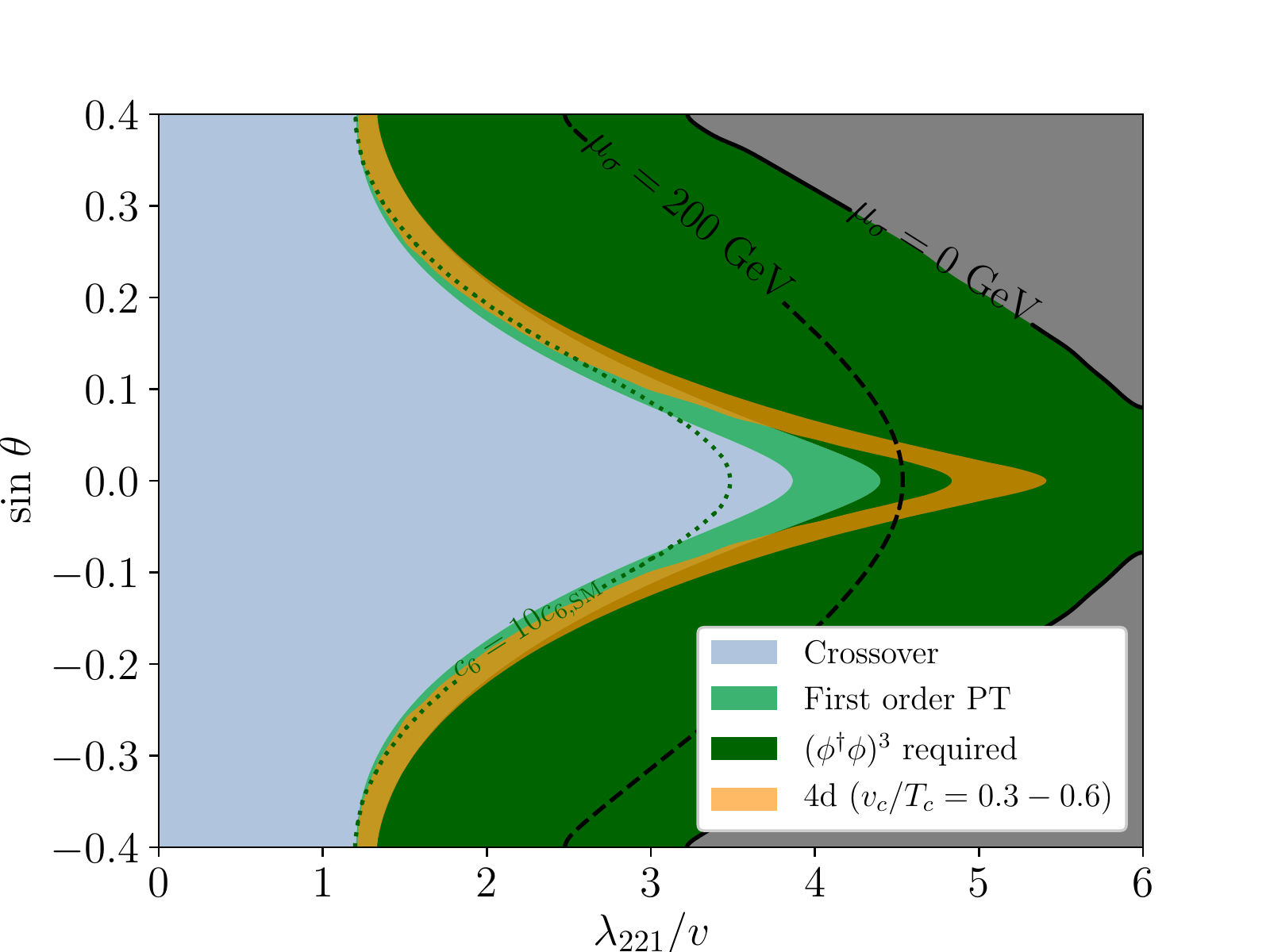}}
    
  \end{center}

  \caption{\label{fig:boomerang_compare_3-d_and_4-d} Parameter space of the xSM predicting a first-order electroweak phase transition for various values of $m_2$ and $\sin \theta$ with $b_3=0$, in both the 4-d perturbative and 3-d nonperturbative approaches. The light green regions features a first-order electroweak phase transition as predicted by existing lattice results. The darker shaded region features $x<0$ so that the higher dimension operator $(\phi^\dagger \phi)^3$ must be kept in the dimensionally reduced theory to resolve one-step electroweak transitions. Furthermore, to the right of the dotted green contour, the dimension-6 operator coefficient, $c_6$, becomes larger than 10 times the corresponding SM value, $c_{6, {\rm SM}}$, and the neglected  $(\phi^\dagger \phi)^3$ term can already have significant effects. We also show contours indicating the size of the singlet \MSbar mass parameter $\mu_\sigma$. When $\mu_\sigma \lessapprox g T \approx 100$ GeV, the superheavy singlet approximation becomes compromised. The gray shaded regions correspond to where the singlet mass parameter $\mu^2_{\sigma}<0$ so that the superheavy dimensional reduction completely breaks down. The parameter space in which 4-d perturbation theory predicts a first-order transition with $v_c/T_c = 0.3-0.6$ is shaded orange, and matches up well with the nonperturbative first-order regions especially in regions where the $(\phi^\dagger \phi)^3$ contribution is unimportant.}

\end{figure*}

We now move on to the more general case without a $Z_2$ symmetry. We first fix $b_3=0$ and vary $\sin \theta$ and $\lambda_{221}$. Our results are shown in Fig.~\ref{fig:boomerang_compare_3-d_and_4-d} for $m_2=240$ GeV and $m_2=400$ GeV. In the light green regions our analysis predicts a first-order electroweak phase transition, while the darker green region features $x<0$, and therefore require the $(\phi^\dagger \phi)^3$ operator in the 3-d EFT to explore one-step phase transitions. We also show where $c_6=10 c_{6,{\rm SM}}$, assuming $T=140$ GeV. To the right of the dotted green contour, singlet-induced dimension-6 effects can already be non-negligible. The gray region features $\mu^2_{\sigma}<0$ for which the superheavy dimensional reduction completely breaks down. Also shown is a contour indicating where $\mu_{\sigma}=200$ GeV. In the $m_2=240$ GeV case the superheavy approximation can lead to inaccuracies in determining the first-order phase transition parameter space. For $m_2=400$ GeV the superheavy approximation is better justified, and we expect our nonperturbative analysis to be more accurate.  
 Corresponding results for the parameter space with varying $b_3$ are shown for $m_2=240$, 400 GeV and $\sin \theta = 0.05$, 0.2 in Fig.~\ref{fig:L221_mu3}. In these slices, portions of the gray shaded regions are excluded by 1-loop absolute vacuum stability, as discussed in Ref.~\cite{Chen:2017qcz} (see also Ref.~\cite{Gonderinger:2009jp}).

\begin{figure*}[!t]

  \subfigure[\; $m_2 = 240$ GeV, $b_4 = 0.25$, sin$\theta = 0.05$]{
    \includegraphics[width=0.4\textwidth]{%
      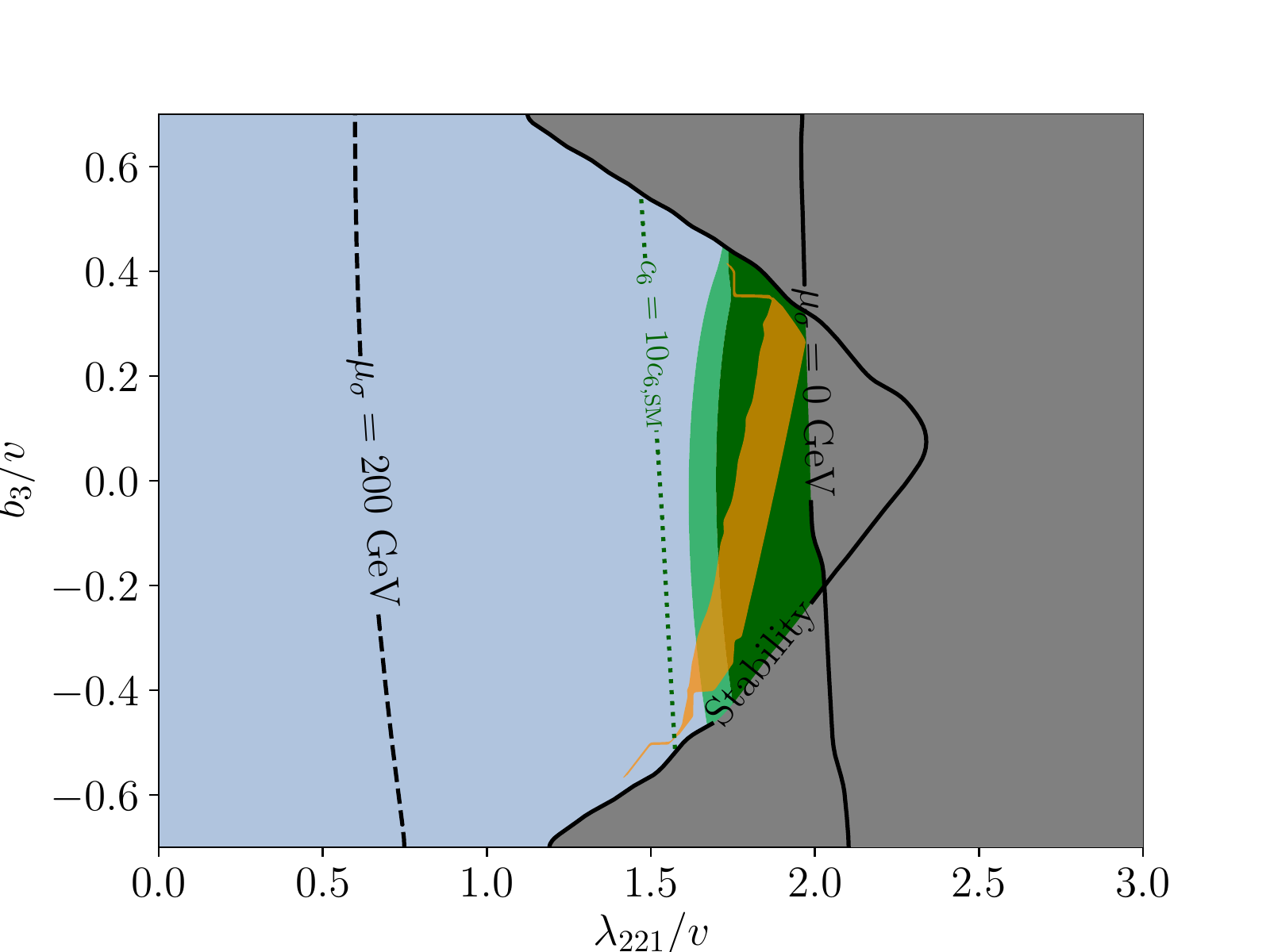}}
  \,
  \subfigure[\; $m_2 = 240$ GeV, $b_4 = 0.25$, sin$\theta = 0.2$]{
    \includegraphics[width=0.4\textwidth]{%
      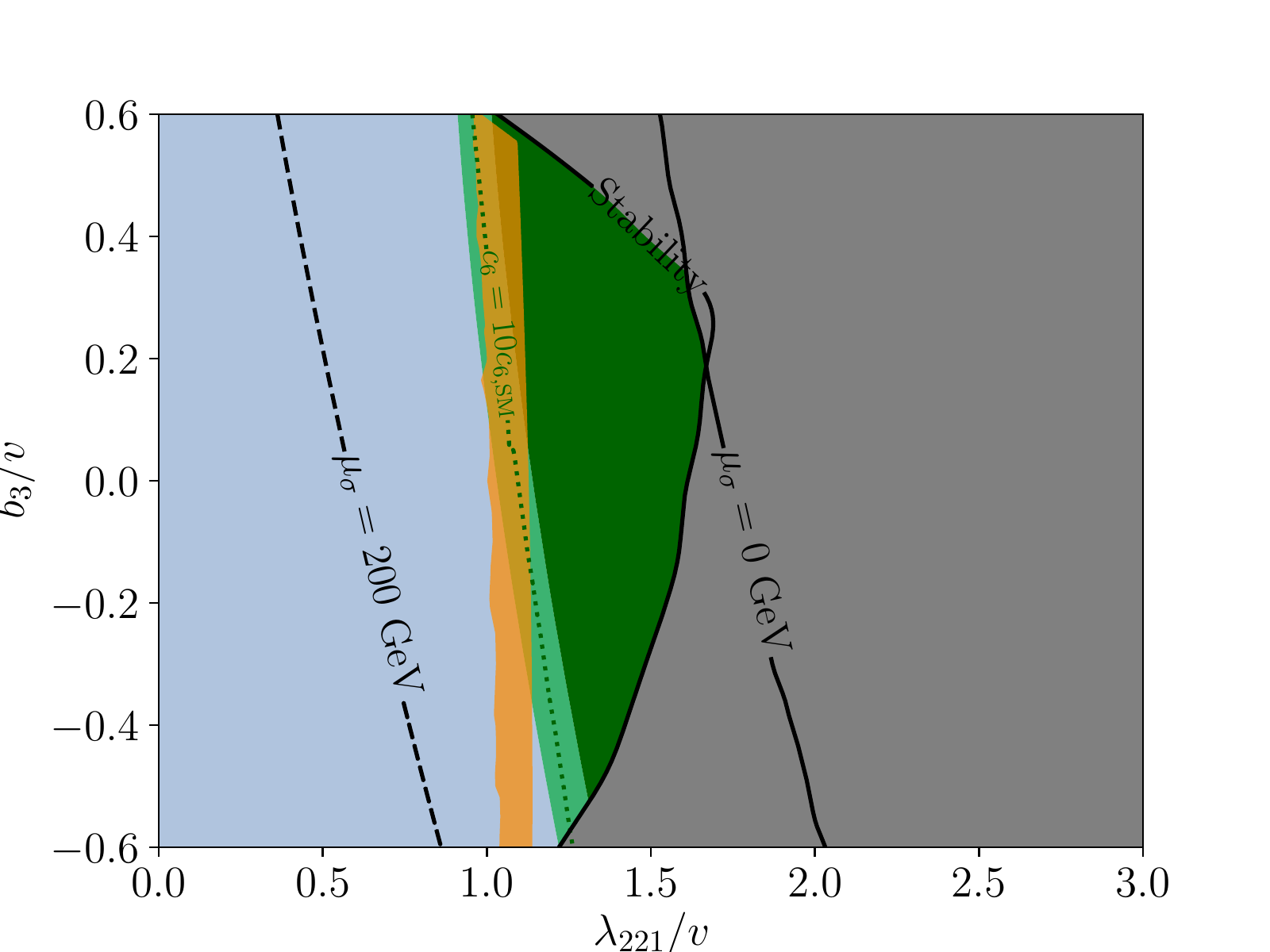}}
  \\
  \subfigure[\; $m_2 = 400$ GeV, $b_4 = 0.25$, sin$\theta = 0.05$]{
    \includegraphics[width=0.4\textwidth]{%
      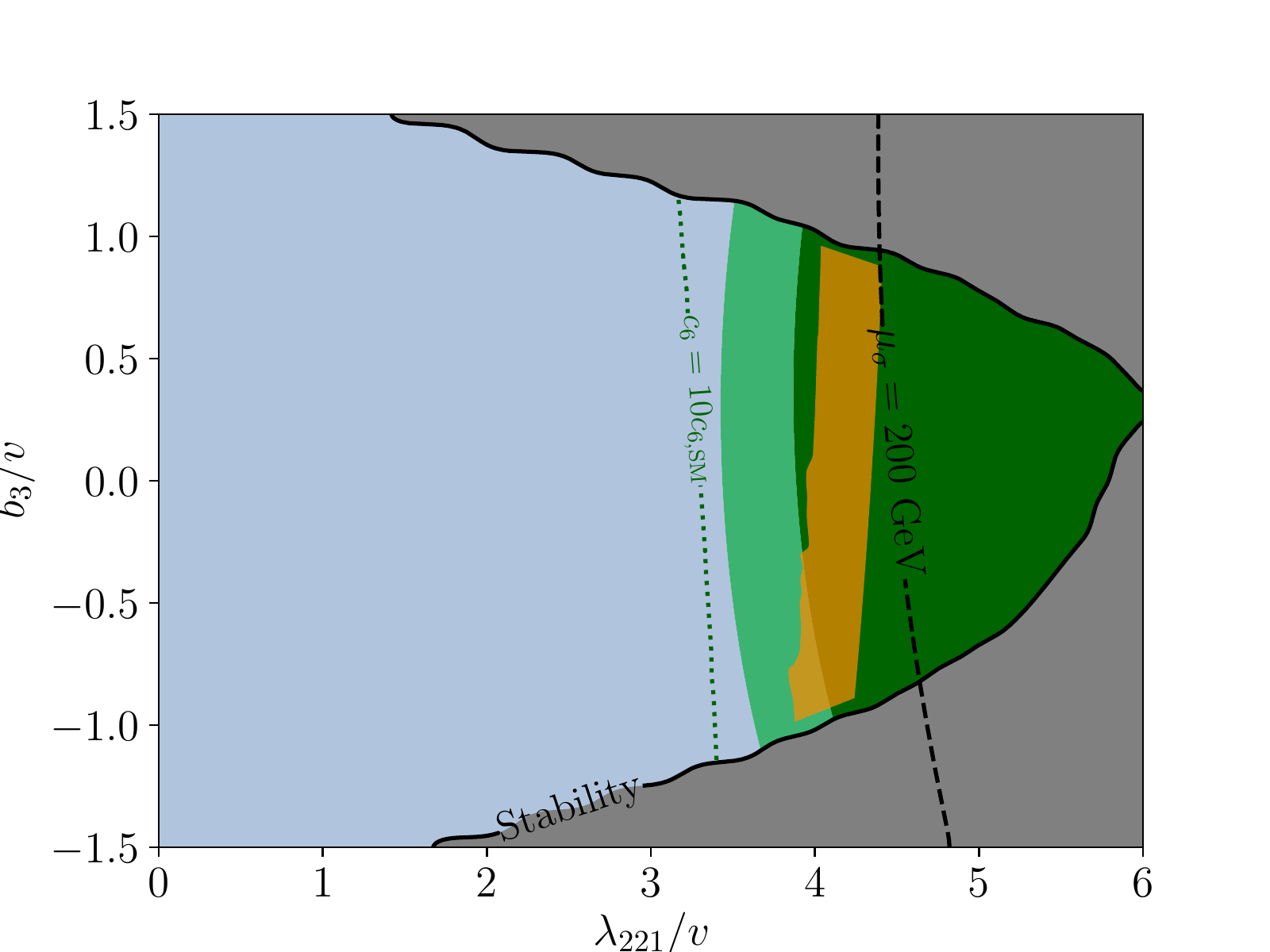}}
  \,
  \subfigure[\; $m_2 = 400$ GeV, $b_4 = 0.25$, sin$\theta = 0.2$]{
    \includegraphics[width=0.4\textwidth]{%
      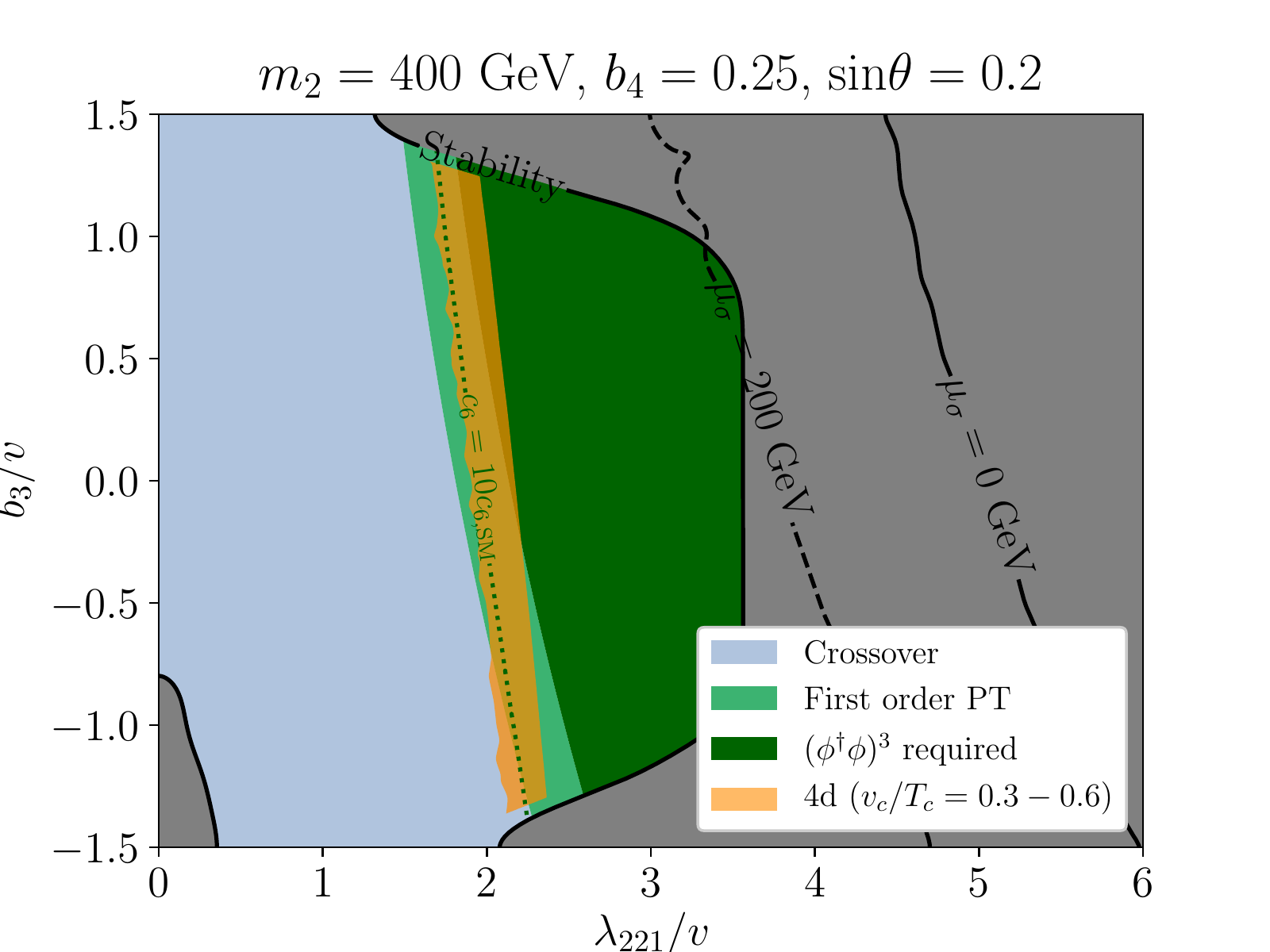}}

\caption{Parameter space of the xSM predicting a first-order phase transition for various values of $m_2$ and $\sin \theta$ away from the $b_3=0$ limit. The shading and contours are as in Figs.~\ref{fig:Z2}-\ref{fig:boomerang_compare_3-d_and_4-d}, with additional contours and gray shaded regions indicating where the electroweak vacuum is metastable in a 1-loop analysis. Our nonperturbative predictions for the first-order electroweak phase transition regions correspond roughly to the parameter space in which 4-d perturbation theory predict a first-order transition with $v_c/T_c = 0.3-0.6$. \label{fig:L221_mu3}}

\end{figure*}

As can be seen from Figs.~\ref{fig:Z2}-\ref{fig:L221_mu3}, the neglected singlet-induced dimension-6 operator can become important in the first-order transition regions, especially for small $|\sin\theta|$ and in the $Z_2$ limit. This is expected from simple power counting arguments. Comparing Eq.~(\ref{eq:lam3}) and Eq.~(\ref{eq:c6}) for the approximate dimension-6 operator coefficient, we see that generally, for superheavy singlets such that $\mu_\sigma \sim \pi T$, the effect of the singlet on the dimension-6 operator is suppressed by $\sim a_2/\pi^2$ relative to the leading effect on $\lambda_3$ (and hence the strength of the phase transition). However, for $|\sin \theta| \approx 0$, the effect on $\lambda_{h,3}$ starts at $\mathcal{O}(a_2^2/16 \pi^2)$, rather than at  $\mathcal{O}(g^2/8 \pi^2)$ for $a_1\sim g T$, and larger values of $a_2$ are required to achieve a given value of $x$. Thus, for small $|\sin \theta|$, the $ a_2/\pi^2$ suppression of the dimension-6 operator is reduced, rendering these contributions more important. Our results motivate new nonperturbative studies including these dimension-6 operators, which will be required to more accurately study first-order transitions in the small $|\sin \theta|$ and $Z_2$ regions. As discussed in Sec.~\ref{sec:pheno} below, these are precisely the regions requiring future colliders to experimentally probe.

\subsection{Comparison with perturbation theory}

It is useful to compare the results above to those obtained by the conventional perturbative approach. In a perturbative setting, one typically analyzes the phase structure of the theory using the finite temperature effective potential, $V_{\rm eff}$. To one loop order, and in a mass-independent renormalization scheme, it is given by
\begin{equation}\begin{aligned}
V_{\rm eff}= &V_0 + \sum_i \frac{\pm n_i}{64 \pi^2} m_i^4(h,\sigma)\left[\log\left(\frac{m_i^2(h,\sigma)}{\mu_R^2}\right) - c_i \right] \\
&+ \frac{T^4}{2\pi^2} \sum_i \pm n_i J_{\pm}\left(\frac{m_i^2(h,\sigma)}{T^2}\right).
\end{aligned}
\end{equation}
Here the sums run over all species coupled to the scalar fields, with upper (lower) signs for bosons (fermions), $n_i$ are the number of degrees of freedom for the species $i$, $c_i$ are scheme-dependent constants, $\mu_R$ is the renormalization scale, and the $J_\pm$ encode the thermal corrections (see e.g. Ref.~\cite{Quiros:1999jp} for full expressions). To improve the perturbative expansion away from the symmetric phase, one can insert thermally-corrected masses for the $m_i$ above. This procedure resums so-called daisy diagrams and delays the infrared breakdown of perturbation theory to smaller field values. Typically a high-temperature expansion is made for the thermal self-energies in this step, although there has been recent progress in going beyond this approximation~\cite{Curtin:2016urg}. In what follows, we use the high-$T$ thermal masses, the expressions for which can be found in e.g.~Refs.~\cite{Chen:2017qcz}, as this is the most conventional approach. The effective potential is computed in Landau gauge utilizing the \MSbar scheme.

With $V_{\rm eff}$ defined above, we search for first-order electroweak phase transitions in the same way as that detailed in Ref.~\cite{Chen:2017qcz}, tracing the minima of the finite-temperature effective potential up in temperature until the electroweak minimum is degenerate with a vacuum with restored electroweak symmetry. This defines the critical temperature, $T_c$, and critical field value, $v_c$. It is important to note that this approach yields gauge-dependent results for the critical temperature and order parameter of the phase transition. Ref.~\cite{Patel:2011th} provides a systematic method for avoiding the spurious gauge dependence in determining $T_c$. It would be interesting to perform a comparison of the results obtained by this method with our (manifestly gauge-invariant) nonperturbative approach in the future. However, in our present study we restrict ourselves to the conventional (gauge-dependent) method described above.
 
 In the perturbative approach, a thermal cubic term is always present in the finite temperature effective potential. This means that perturbation theory always predicts a first-order transition. This is a spurious effect due to the IR breakdown of the perturbative expansion away from the broken phase and prevents one from resolving the endpoint of the transition. Therefore, to extract meaningful information about the phase structure of the theory in this approach, one must restrict themselves to considering sufficiently ``strong'' first-order transitions. The strength of the phase transition can be parametrized by the order parameter $v_c/T_c$. Conventionally, strong first-order transitions are defined by imposing a minimum value for the order parameter, $v_c/T_c \geq \zeta$. The value of $\zeta$ depends on the particular question of interest. In the setting of electroweak baryogenesis, it is determined by requiring sufficient sphaleron suppression in the broken phase, typically corresponding to $\zeta \approx 0.6-2$ (see e.g.~Refs.~\cite{Quiros:1999jp, Patel:2011th} for a more detailed discussion). From the standpoint of observable gravitational waves, $\zeta$ typically needs to be larger, although the strength of the phase transition in this case is usually parameterized by the more physical quantity $\alpha$. For the purpose of benchmarking perturbation theory, we consider values of $\zeta$ that most closely reproduce the first-order transition region accessible by our nonperturbative analysis and in which the various assumptions of Secs.~\ref{sec:NP}-\ref{sec:GW} are justified. This will allow us to compare how the phase structure of the theory depends on the various model parameters in the perturbative and nonperturbative approaches. In practice this amounts to considering
 \begin{equation}
 \zeta \in [0.3, 0.6].
 \end{equation}
 Such transitions are quite weak from the standpoint of sphaleron suppression and gravitational wave generation (as we will see below) but serve to delineate the regions of parameter space in which current nonperturbative studies concretely indicate a genuine first-order electroweak phase transition.

The region with $v_c/T_c \in [0.3, 0.6]$ in 1-loop 4-d perturbation theory  is indicated by the shaded orange region in Fig.~\ref{fig:Z2} for the $Z_2$-symmetric case. The couplings required for a ``strong'' first-order EWPT in the 4-d approach tend to be somewhat larger than those corresponding to first-order EWPTs in our nonperturbative analysis. However, it is important to note that the superheavy approximation is not justified across most of the first-order PT region in Fig.~\ref{fig:Z2}, and that the neglected dimension-6 operators can be important here, as discussed above. Our 4-d treatment retains the singlet field in the effective potential and therefore accounts for the effects corresponding to the higher-dimension operators. Future nonperturbative studies including these effects will be required to conclusively benchmark the predictions from perturbation theory in this limit. 

The orange shaded regions of Figs.~\ref{fig:boomerang_compare_3-d_and_4-d}-\ref{fig:L221_mu3} show the parameter space for which $v_c/T_c \in [0.3, 0.6]$ in the non-$Z_2$ case, where the transition is strengthened mainly by the tree-level contributions from the singlet, cf.~Eq.~(\ref{eq:lam3}). These regions line up quite well with the first-order EWPT parameter space predicted by our nonperturbative analysis, especially where the $(\phi^\dagger \phi)^3$ effects are expected to be small. The parametric dependence of the strong first-order region predicted by 4-d perturbation theory reliably traces that of the first-order EWPT regions obtained by our nonperturbative analysis in portions of the parameter space where our nonperturbative treatment is well-justified. Our results suggest that, in the particular 4-d approach considered, a genuine first-order EWPT tends to correspond roughly to $v_c/T_c\gtrsim 0.3$. It should be emphasized that this conclusion is gauge-dependent and not necessarily true for other gauge choices.

\section{Exploring the phase diagram at gravitational wave and collider experiments}\label{sec:pheno}

Having determined the singlet model parameter space where the electroweak phase transition goes from a thermodynamic cross-over to a first-order transition, we would like to consider the prospects for studying the nature of the electroweak phase transition experimentally in this scenario. Below, we discuss gravitational wave and collider probes of the first-order EWPT parameter space obtained from the nonperturbative methods described in Sec.~\ref{sec:GW}. While the transitions accessible by the repurposed nonperturbative results are quite weak and therefore challenging to detect at experiments like LISA, it is nevertheless informative to compare the results from the 3-d and 4-d approaches. This serves as a benchmark of perturbation theory, and indicates the levels of uncertainty inherent in perturbative estimates of the resulting gravitational wave signal.
 
\subsection{Benchmarking for gravitational wave predictions}
\label{sec:ssb_gws}

\begin{table}
\label{tab:sample-point}
  \begin{center}
      \begin{ruledtabular}
        \begin{tabular}{cdddd}
          &
          \multicolumn{1}{c}{ $T_c$/GeV }&
          \multicolumn{1}{c}{ $T_n$/GeV }&
          \multicolumn{1}{c}{$\alpha(T_c)$}&
          \multicolumn{1}{c}{ $\beta/H_*$}  \\
          \colrule
          NP     &140.4 &140.2 &0.011 &8.20 \times 10^4 \\
          3-d PT &140.4 &140.0 &0.010 &6.11 \times 10^4 \\
          4-d PT &131.0 &130.7 &0.004 &5.59 \times 10^4 \\
        \end{tabular}
      \end{ruledtabular}
  \end{center}

  \caption{Sample of thermodynamic quantities relevant for
    determination of gravitational wave power spectrum. These values
    correspond to the xSM input parameters $m_2 = 400$ GeV, $b_3=0$
    GeV, $b_4 = 0.25$, $\sin(\theta)=0.2$, $\lambda_{221}/v=2.22$. For
    these inputs the parameter $\lambda_{221}/v$ is tuned such that at
    the critical point $x_c = 0.036$. This allows us to repurpose the
    only pre-existing nonperturbative result for bubble nucleation.
    In Appendix~\ref{sec:DR-accuracy} we estimate that the uncertainty
    in the quantities in the first row to be a few tens of
    percent. \label{tab:GW_input}}

\end{table}

For a nonperturbative determination of the gravitational wave power spectrum in the xSM, we follow the general recipe outlined in Section \ref{sec:gw_implications}. We show the values of $T_c$, $T_n$, $\alpha(T_c)$ and $\beta/H_*$ calculated using our nonperturbative approach (NP) for a representative sample point in Table \ref{tab:GW_input}.
Also shown are the same quantities calculated using the usual 4-d perturbative approach (4-d PT) -- see Section~\ref{sec:identifying_fopt} -- as well as our two-loop, fully-perturbative approach utilising the DR (3-d PT) -- see Section \ref{sec:gw_implications}.

A comparison of the nonperturbative and perturbative results in Table \ref{tab:GW_input} shows agreement on the order of magnitude of all quantities. This provides an important validation for the usual 4-d perturbative methods, at least in this region of parameter space, and a benchmark of the numerical importance of underlying uncontrolled approximations~\cite{Linde:1980ts,Gross:1980br,Weinberg:1992ds,Gleiser:1993hf,Alford:1993br,Surig:1997ne,Garbrecht:2015yza,Metaxas:1995ab,Baacke:1999sc,Patel:2011th,Garny:2012cg,Strumia:1999fv}. There is nevertheless a 10\% discrepancy in $T_c$, a factor of 3 discrepancy in $\alpha(T_c)$ and a 30\% discrepancy in $\beta/H_*$. However, the ratio $T_n/T_c$ agrees very well, to better than 1\%. We have verified that similar trends apply to other points as well in the relevant parts of the xSM parameter space.

Due to the Linde problem (the breakdown of perturbation theory), it is not possible to make a fully reliable error estimate within perturbative theory. For our nonperturbative approach however, this is possible, and in Appendix \ref{sec:DR-accuracy} we discuss the various sources of uncertainty and their expected sizes. The main sources of uncertainty in the nonperturbative calculation are due to renormalization scale dependence and the neglect of dimension-6 operators in the dimensionally reduced theory~\cite{Niemi:2018asa}. The conclusion of the analysis in Appendix \ref{sec:DR-accuracy} is that we expect our uncertainties in $T_c$, $\alpha(T_c)$ and $\beta/H_*$ to be of order 1\%, 40\% and 60\% respectively. The dominant uncertainty is due to the neglect of dimension-6 operators, which get relatively large corrections from singlet-Higgs interactions, in particular from $a_1$ (for reference, $c_6 \approx 16 c_{6, {\rm SM}}$ for this benchmark point). From the 4-d perturbative standpoint, there are significant uncertainties due to the residual renormalization scale dependence of our one-loop approach. For the values reflected in Table~\ref{tab:GW_input} the renormalization scale was set to $\mu_R = (m_1+m_2)/2$. Varying $\mu_R$ between $m_Z$ and $2m_2$, we find uncertainties of $\sim5\%$ in $T_c$, $T_n$, $\sim 50\%$ in $\alpha(T_c)$, and $\sim 80\%$ in $\beta/H_*$. Taking this into account we conclude that, except for $T_c$, the perturbative and nonperturbative approaches agree within uncertainties.   Note that at smaller values of $x$, corrections from dimension-6 operators will be comparatively larger, hence we expect that one cannot trust our DR for the xSM at $x$ significantly smaller than $x=0.036$, reflected in Table~\ref{tab:GW_input}.

From Table \ref{tab:GW_input} and the fit formula of Ref.~\cite{Hindmarsh:2017gnf} one can deduce that, for this benchmark point in the xSM, the peak frequency of the gravitational wave signal today is of order 1~Hz. This is near the peak sensitivity of both DECIGO~\cite{Kawamura:2011zz} and BBO~\cite{Harry:2006fi} though the strength of the signal is too weak to be detected. We plot this point in Fig.~\ref{fig:GWs_b}, along with the predictions from 3-d and 4-d perturbation theory. The sensitivity curves shown assume $v_w=1$. A perturbative calculation of the wall velocity for similar points in the xSM was performed in Ref.~\cite{Kozaczuk:2015owa}, which found $v_w\sim 0.1$ for relatively weak transitions such as those we consider here. Assuming this value only weakens the gravitational wave signal, and so we have shown projections for $v_w=1$ to remain conservative in our conclusions. Our scan over the xSM parameter space finds that $\eta_y(T_c)$ lies in the range [3.85,5.45] almost independently of $x$. This region is shaded dark blue on the right in in Fig.~\ref{fig:GWs_b}, indicating values of $\alpha$ and $\beta/H_*$ in the parameter space where the singlet model is described by the 3-d SM-like EFT in the infrared. This region again lies outside that detectable by planned gravitational wave experiments.

For a stronger transition there are two options: either the BSM field must play an active, dynamical role or the dimension-6 operators in the 3-d EFT must be included. In the case of the xSM, this is in agreement with previous perturbative studies (see e.g.~Ref.~\cite{Alves:2018jsw}).

\subsection{Implications for colliders}

\begin{figure*}

  \begin{center}

    \subfigure[\; $b_4 = 0.25$, $b_3 = 0 $, $a_1 = 0 $]{
      \includegraphics[width=0.45\textwidth]{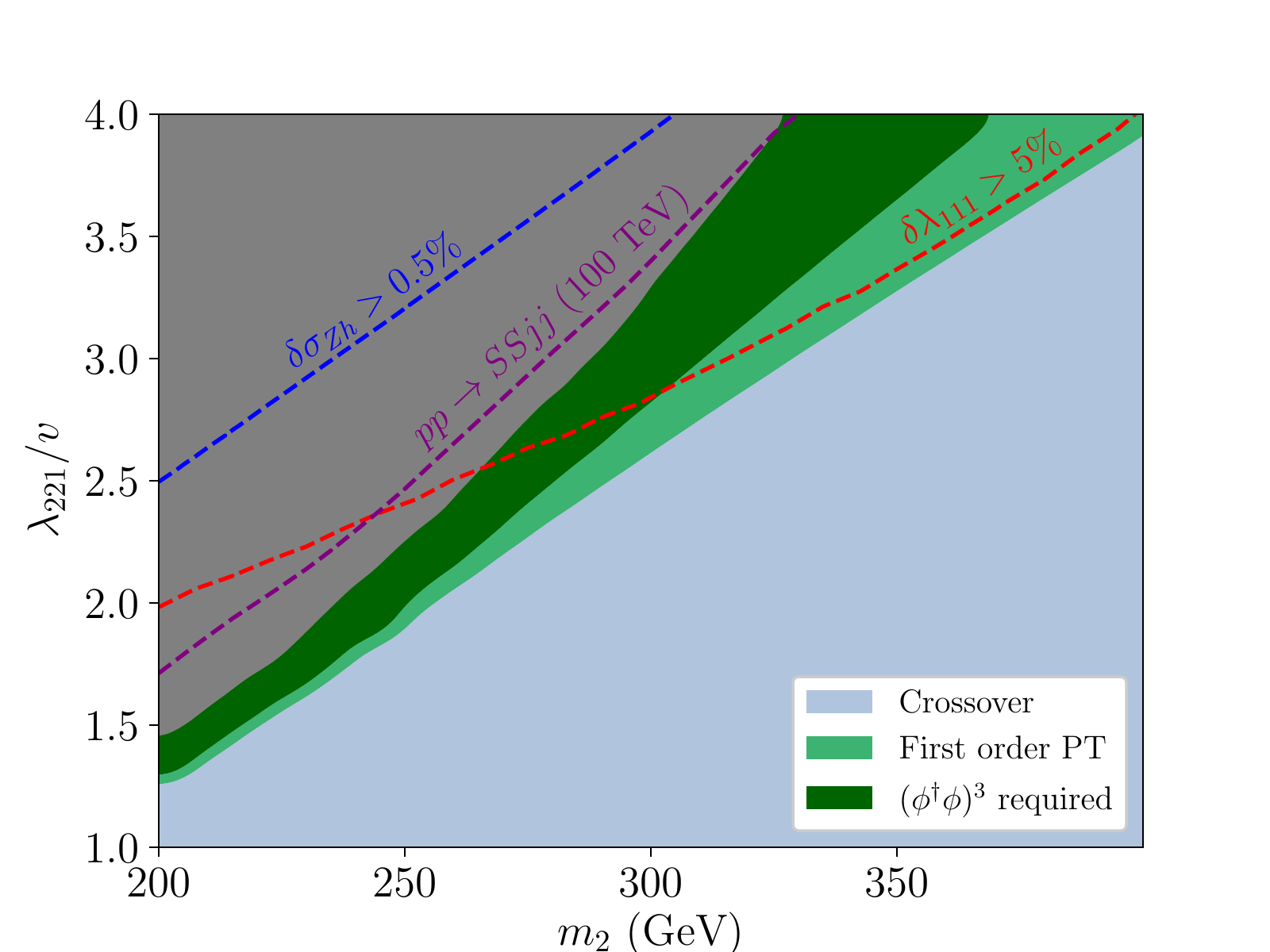}}
    \subfigure[\; $m_2 = 400$ GeV, $b_4 = 0.25$, $b_3 = 0$]{
      \includegraphics[width=0.45\textwidth]{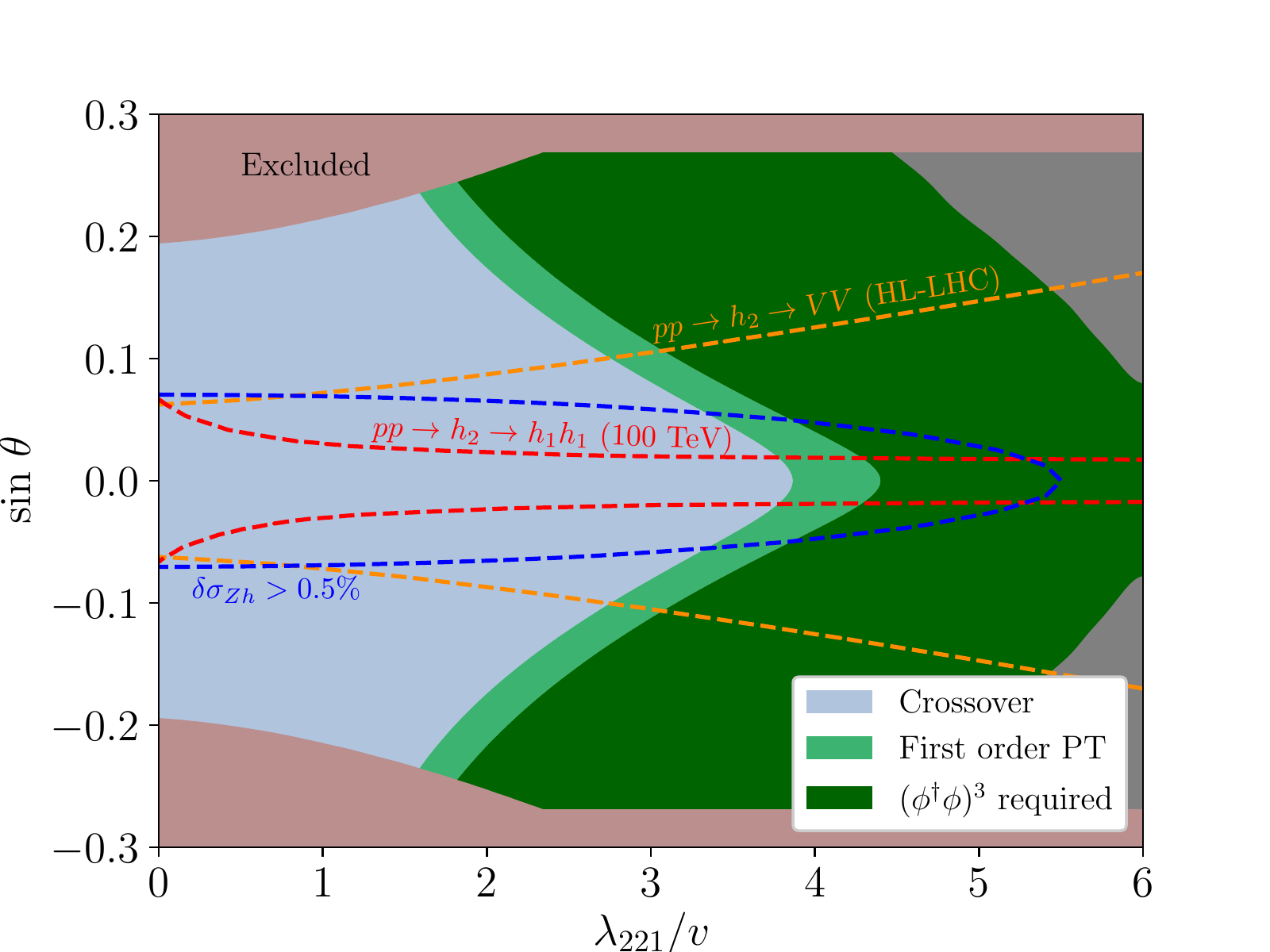}}
    
  \end{center}

  \caption{\label{fig:pheno} As in Fig.~\ref{fig:Z2} ($Z_2$) and the right panel of \ref{fig:boomerang_compare_3-d_and_4-d} (non-$Z_2$) but including limits and projections for several collider observables, discussed in the text. The red shaded regions on the right are currently excluded either by the electroweak precision measurements discussed in Ref.~\cite{Robens:2015gla}, or direct searches for the singlet-like scalar in final states involving two gauge bosons~\cite{Aaboud:2017rel}. The high luminosity LHC, a future Higgs factory and 100 TeV collider could together probe a significant amount of parameter space consistent with a first-order electroweak phase transition as predicted by our nonperturbative analysis. }

\end{figure*}

While gravitational wave experiments such as LISA will likely be able to probe strong electroweak transitions in this model (for which nonperturbative studies do not exist), the transitional regions between a cross-over and a first-order EWPT can still be interesting cosmologically and will likely require other probes to access. Fortunately, colliders can be sensitive to precisely these regions and complement gravitational wave experiments in exploring the singlet model electroweak phase diagram. There has been a significant amount of work on the collider phenomenology of singlet models over the years, and a large number of possible experimental signatures of the new scalar have been proposed~\cite{Profumo:2007wc, Noble:2007kk, Assamagan:2016azc}. Some of these are directly related to $\lambda_{221}$ that is one of the couplings correlated with the nature of the electroweak phase transition in both the perturbative and nonperturbative approaches. Two such observables are the $h_2 h_2$ production cross-section at hadron colliders~\cite{Craig:2014lda, Curtin:2014jma, Chen:2017qcz}, and the shift in the Higgs couplings from their SM predicted values~\cite{Profumo:2007wc, Katz:2014bha, Profumo:2014opa}, most notably the $h_1ZZ$ coupling~\cite{Curtin:2014jma, Huang:2016cjm}. Complementary information about the scalar potential can be inferred from di-Higgs processes~\cite{Noble:2007kk, Huang:2015tdv, Kotwal:2016tex, Huang:2016cjm, Huang:2017jws} as well as direct production of the singlet and subsequent decays to SM gauge bosons or fermions~\cite{Buttazzo:2015bka} at hadron colliders.

We demonstrate the interplay between some of the aforementioned experimental signatures on the parameter space predicting a first-order electroweak phase transition in Fig.~\ref{fig:pheno}. The left panel shows collider projections for the $Z_2$-symmetric case. This model is very difficult to probe at the LHC but provides a compelling target for future colliders~\cite{Craig:2014lda, Curtin:2014jma}. An $e^+e^-$ collider operating as a Higgs factory is expected to achieve excellent precision in measuring the $Zh$ cross-section, $\sigma_{Zh}$ (where $h$ is the SM-like Higgs). The parameter space lying above the blue dashed contour feature $\delta \sigma_{Zh}>0.5\%$, computed at
one-loop using the expressions found in Ref.~\cite{Chen:2017qcz}. This corresponds approximately to the expected sensitivity of future circular Higgs factories such as the FCC-ee, or CEPC to this observable. Regions above the purple dashed contour could be probed at the 95\% confidence level by searches for $p p \rightarrow h_2 h_2 j j$ at a 100 TeV collider with 30 ab$^{-1}$ integrated luminosity. These projections are taken from Ref.~\cite{Curtin:2016urg}, corrected to account for our slightly different choice of renormalization scale. While this search will likely not be sensitive to the EWPT parameter space accessible by our nonperturbative analysis, it should be able to probe stronger two-step phase transitions, which will require new lattice studies to explore nonperturbatively.  The region above the red dashed contour features a deviation in the Higgs self-coupling, $\delta \lambda_{111} \equiv |\lambda_{111} - \lambda_{111, {\rm SM}}|/ \lambda_{111, {\rm SM}}$ greater than 5\%, which corresponds roughly to the level of sensitivity expected to be achieved at a future 100 TeV collider with 30 ab$^{-1}$~\cite{Benedikt:2018csr}. Such measurements will be able to probe the region for which existing nonperturbative studies predict a first-order phase transition, although it is likely that including the relevant higher-dimension operators will be required to improve the nonperturbative predictions.

Similar results for the non-$Z_2$ parameter space are shown for $m_2=400$ GeV,  $b_3=0$ on the RHS of Fig.~\ref{fig:pheno}, where we show current constraints on the parameter space: the red shaded regions are excluded either by electroweak precision measurements discussed in Ref.~\cite{Robens:2015gla}, or direct searches for the singlet-like scalar in final states involving two gauge bosons. In deriving the latter, we use the 13 TeV ATLAS search in Ref.~\cite{Aaboud:2017rel} and the narrow width approximation, with the $h_2$ production cross-section taken from Refs.~\cite{Dittmaier:2011ti, Heinemeyer:2013tqa} for a SM-like Higgs of the same mass, rescaled by $\sin^2\theta$. We also show projected sensitivities for the same searches ($pp\rightarrow h_2 \rightarrow VV$) at the 13 TeV HL-LHC with 3 ab$^{-1}$ (orange dashed contours), obtained from a simple rescaling of the sensitivity in Ref.~\cite{Aaboud:2017rel}. The parameter space outside the these sets of contours would be probed at the 95\% confidence level and include a sizable portion of the first-order transition region.

We also show projections for future colliders in Fig.~\ref{fig:pheno}. The blue dashed contours indicate regions for which $\delta \sigma_{Zh}>0.5\%$ (here again $h$ is understood to refer to the SM-like Higgs, $h_1$ in our case). The parameter space outside of these contours is expected to be probed by future Higgs factories.  Also shown in Fig.~\ref{fig:pheno} is the approximate sensitivity of a 100 TeV $pp$ collider to the first-order EWPT parameter space through (resonant) di-Higgs production ($pp\rightarrow h_2 \rightarrow h_1 h_1$). Ref.~\cite{Kotwal:2016tex} suggests that a 100 TeV collider with 30 ab$^{-1}$ integrated luminosity will be able to probe $\sigma \times$BR$(pp\rightarrow h_2 \rightarrow h_1 h_1)$ down to $\sim 100$ fb at 95\% C.L.~for $m_2\approx 400$ GeV. The projections shown assume this reach, and make use of the leading order gluon fusion $h_2$ production cross-section in the narrow width approximation, multiplied by a $k$-factor of 2.3 to account for higher-order corrections (this $k$-factor was chosen to yield the gluon fusion cross-section for a 125 SM-like Higgs at 100 TeV reported in Ref.~\cite{HiggsxS}). The 100 TeV collider reach is quite impressive, extending down to very small mixing angles in the parameter space shown.

From Fig.~\ref{fig:pheno} it is clear that the LHC and future colliders will be able to access much of the parameter space predicting a genuine first-order EWPT in this model. The most difficult to access region is that featuring very small $|\sin\theta|$. However, this region is tuned (away from the $Z_2$ limit) and could in principle be probed if a future $e^+ e^-$ collider were able to measure the $Zh$ cross-section to 0.25\% precision. Given the different parametric dependence of the cross-sections for $pp\rightarrow h_2\rightarrow VV, h_1 h_1$ and $e^+ e^- \rightarrow Zh_1$, it will be important to perform all of these different measurements, if possible, in order to gain clear insight into the nature of the electroweak phase transition. Similar conclusions hold for different slices of the model parameter space. From the results shown in Fig.~\ref{fig:pheno} it is clear that colliders will play an important role in exploring the phase structure in this model and complement the reach of gravitational wave experiments in this regard. The regions of parameter space most challenging to access for colliders will require new simulations including the effect of the higher dimension operators to more accurately study nonperturbatively.

\section{Conclusions}
\label{sec:conclusions}

In this work, we have
demonstrated the full pipeline for obtaining theoretically sound, nonperturbative results for the gravitational wave power spectrum arising from a first-order phase transition in the early Universe. Our proof of principle calculation collects and utilizes several tools developed in the literature for the SM with a light Higgs,
and repurposes pre-existing lattice results in the dimensionally reduced effective theory for equilibrium \cite{Kajantie:1995kf} and non-equilibrium thermodynamic quantities \cite{Moore:2000jw}. These relatively few quantities are then used as input parameters for the analysis of numerical simulations of the coupled
fluid-scalar field system \cite{Hindmarsh:2017gnf} in order to determine the prediction for the gravitational wave power spectrum from a first-order electroweak phase transition in general extensions of the Standard Model matching on to the infrared 3-d SM-like EFT.

The physics of the phase transition is largely determined by the long-distance 3-d effective field theory.
 Our analysis suggests that any SM extension which reduces to the SM-like 3-d EFT will have a rather weak EWPT and therefore be unobservable at gravitational wave experiments in the foreseeable future. To accommodate a strong transition producing an observably large gravitational wave signal, either new light fields or sizable dimension-6 operators must turn up in the IR effective theory. Otherwise, collider experiments are likely to provide the best chance of probing first-order electroweak phase transitions in these scenarios, which can still be interesting cosmologically despite the small corresponding gravitational wave signal.

We have applied these results to the real singlet extension of the Standard Model, identifying the regions where the SM-like electroweak crossover becomes a first-order transition. In the regions we have been able to analyze by repurposing pre-existing nonperturbative results -- where the singlet field plays a non-dynamical role and induces negligible effects through higher dimension operators -- the gravitational wave signal is unfortunately far too weak to be detected with LISA or other near-future gravitational wave experiments, despite the fact that the electroweak phase transition can be of first-order. On the other hand, a combination of measurements at the high luminosity LHC and future colliders can in fact probe much of the first-order transition regions in the singlet model. Our nonperturbative results provide new targets for exploring the phase structure of the theory at colliders that cannot be delineated in a perturbative approach.

Observable gravitational wave signals in the xSM require the singlet field to play a dynamical role in the transition or induce non-negligible effects through higher dimension operators. This in turn makes the demand for new nonpertubative simulations even more urgent. Nevertheless, using exiting results we are able to benchmark the predictions of perturbation theory for the phase transition parameters relevant for computing the gravitational wave power spectrum. We find that perturbative results reproduce the latent heat and duration of the phase transition to within $\mathcal{O}(1)$ factors, and the nucleation temperature to within 10\% of the nonperturbative prediction, agreeing to within the expected uncertainties in our approach. It will be worthwhile and interesting to perform the analogous comparison for points within the LISA sensitivity band once new nonperturbative simulations including higher dimension operators or the dynamical singlet field become available.

\section*{Acknowledgments}

The authors would like to thank Daniel Cutting, Patrick Draper, Mark
Hindmarsh, Stephan Huber, Venus Keus, Mikko Laine, Jonathan Manuel,
Jose M. No, Hiren Patel, Kari Rummukainen, Anders Tranberg and Aleksi
Vuorinen for discussions.  The work of JK was supported by NSF grant
PHY-1719642. MJRM was supported in part under U.S. Department of
Energy contract DE-SC0011095. LN was supported by Academy of Finland
grant no. 308791 and the Jenny and Antti Wihuri Foundation.
TT has been supported by the Vilho, Yrj\"{o} and
Kalle V\"{a}is\"{a}l\"{a} Foundation, and by the European Research
Council grant no. 725369.  In addition, TT was supported by the Swiss
National Science Foundation (SNF) under grant 200020-168988.  DJW
(ORCID ID 0000-0001-6986-0517) was supported by Academy of Finland
grant no.~286769. OG and DJW are supported by the Research Funds of
the University of Helsinki. The work of JK, MJRM and DJW was performed
in part at the Aspen Center for Physics, which is supported by
National Science Foundation grant PHY-1607611.

\appendix
\begin{widetext}

\section{Details of relating renormalized parameters to physical observables}
\label{sec:vacuum_renormalization}

As described briefly in Section~\ref{sec:parametrisation}, the input parameters
for our scans of the singlet-extended SM are the pole masses $M_{1}, M_{2}$ (which we have been writing as $m_1$, $m_2$, but in this section are denoted by capital letters to distinguish them from the corresponding \MSbar masses) of the mass eigenstates
$h_1$ and $h_2$, as well as the mixing angle $\theta$ and the couplings
$b_3,b_4,\lambda_{221}$, which are assumed to be inputed directly
at a fixed \MSbar scale. We parameterize the scalar doublet as
\begin{align}
\phi = \begin{pmatrix}
\omega^+ \\
\frac{1}{\sqrt{2}} \left(v + h + i z\right)
\end{pmatrix},
\end{align}
with $\omega^\pm, z$ being the Goldstone modes.

After rotating the scalar potential (\ref{eq:scalarpotential}) to a diagonal basis via
\begin{align}
h_1 &= h\cos\theta - \sigma \sin\theta, \\
h_2 &= h\sin\theta + \sigma \cos\theta,
\end{align}
the mass-eigenvalue relations can be inverted to express the \MSbar parameters in the form
\begin{align}
\label{eq:MSbar-tree}
\mu^2 &= \frac14 \big( m^2_{1} + m^2_{2} + (m^2_{1} - m^2_{2})\cos 2\theta \big), \\
\mu^2_\sigma &= \frac12 \big( m^2_{1} + m^2_{2}  + (m^2_{2} - m^2_{1})\cos 2\theta - v^2 a_2 \big), \\
b_1 &= -\frac14 (m^2_{2} - m^2_{1}) v \sin 2\theta, \\
a_1 &= \frac{m^2_{2} - m^2_{1}}{v} \sin 2\theta, \\
\lambda &= \frac{1}{4v^2} \big( m^2_{1} + m^2_{2} + (m^2_{1} - m^2_{2} )\cos 2\theta \big).
\end{align}
Masses in the diagonal basis have been denoted by $m_{1}$ and $m_{2}$ to distinguish them from the pole masses $M_{1}, M_{2}$. The portal coupling $a_2$ can be solved in favor of the trilinear $(h_2,h_2,h_1)$ coupling of the diagonalized theory via
\begin{align}
\label{eq:lambda_m_tree}
a_2 &= \frac{(m^2_{1} + 2m^2_{2})(\cos 2\theta - 1) + 2 v \lambda_{221} \sec\theta + 2 v b_3 \sin 2\theta}{v^2 (3\cos 2\theta - 1)}.
\end{align}
Inverted relations for the top Yukawa $y_t$ and gauge couplings ${g,g'}$ remain unchanged from the Standard Model case.

We follow Refs.~\cite{Kajantie:1995dw,Laine:2017hdk,Niemi:2018asa} in relating the parameters appearing in the Lagrangian to physical pole masses. In practice this procedure consists of solving for the running masses from loop-corrected pole conditions of the form
\begin{align}
\big[p^2 - m^2 (\Lambda) + \Pi(p^2, \Lambda) \big]_{p^2 = M^2} = 0,
\end{align}
where $\Pi$ denotes (the real part of) a self-energy calculated at one-loop level. In order to reduce the effect of logarithmic corrections, we calculate the self energies with the RG scale $\Lambda$ set equal to $(M_{1}+M_{2})/2$. The pole-mass conditions fix the renormalized masses of $h_1,h_2$ and gauge bosons $W,Z$ at the input scale. The numerical values used for the pole masses are $M_{1} = 125.09$ GeV,corresponding to the observed Higgs boson, $M_W = 80.385$ GeV and $M_Z = 91.188$ GeV. For fermions, the pole condition depends on the scalar and vector parts of the self-energy, and the relevant equation reads
\begin{align}
m_f^2 = M_f^2 (1 - 2 \Sigma_s(M_f^2,\Lambda) - 2 \Sigma_v(M_f^2,\Lambda)).
\end{align}
In practice, we only solve the correction to the top quark mass using $M_t = 173.1$ GeV and neglect the procedure for other fermions as their Yukawa couplings have little effect on the phase transition. However, all fermions are included in the loop corrections.

The correction to the Higgs VEV is calculated by relating it to the electromagnetic fine structure constant $\hat{\alpha}$, in the \MSbar scheme, and evaluating the one-loop correction to on-shell Thomson scattering. Due to the absence of new charged particles in the singlet extension, this correction is equal to its SM value of \cite{Erler:1998sy} $\hat{\alpha}^{-1}(M_Z) \approx 128$, which we input into our parameterization.

The self-energies are renormalized by the counterterms that were previously used in Ref.~\cite{Brauner:2016fla}. Note, however, that in this reference counterterms were listed in Landau gauge, whereas the vacuum renormalization is more convenient to carry out in the Feynman-t'Hooft gauge. Dependence on the gauge parameter is easily added to the counterterms by first evaluating the field renormalization counterterm and requiring the bare parameters to be gauge independent. A one-loop evaluation and renormalization of the self-energies in the singlet-extended SM is straightforward, but since the actual formulae are fairly long, we do not list them here. The self-energies are functions of the renormalized masses and couplings, so for a self-consistent calculation we have to solve a non-linear system of pole-mass equations. The solutions are conveniently found iteratively, and inserting the loop-corrected masses and $\hat{\alpha}$ into the tree-level relations~(\ref{eq:MSbar-tree})-(\ref{eq:lambda_m_tree}) gives the potential parameters at the input scale $\Lambda=(M_{1} + M_{2})/2$ -- and similarly for $y_t,g$ and $g'$.

As a result of this vacuum renormalization procedure, the renormalized parameters that enter the phase transition calculations are shifted by roughly $5\%-30\%$ relative to the values one would obtain by matching to physical observables directly at tree level.
In the parameter scans, the largest corrections occur for the portal couplings $a_1, a_2$ as well as for $\lambda$, which all play an important role in strengthening the phase transition. However, for masses $M_{2} \gtrsim 400$ GeV the loop correction to the singlet mass parameter $\mu^2_\sigma$ can be large enough to cause deviations of order $100\%$ from the tree-level value if additionally $\lambda_{221}/v \gg 1$ or $a_1/v \gg 1$, which may hint at bad convergence of perturbation theory even without considering infrared effects at high temperature. Overall, the effect of these zero-temperature corrections is to shift the phase transition towards smaller couplings and therefore have a non-negligible numerical impact on our EWPT analysis.

\section{Accuracy of the dimensional reduction}
\label{sec:DR-accuracy}

In Ref.~\cite{Brauner:2016fla} dimensional reduction was performed at one-loop
level, leaving out the two-loop contribution to the mass parameter of the doublet
field. This means that our determination of the critical temperature in the xSM is
not as accurate as in the similar studies in the two Higgs doublet model (2HDM)
\cite{Andersen:2017ika, Gorda:2018hvi} or the real-triplet extension of the SM
\cite{Niemi:2018asa}. However, we expect that the character of transition can
still be determined with satisfactory accuracy and regions of first-order
transition can be identified. In fact, the one-loop accuracy of the dimensional
reduction matches the accuracy used in the perturbative four-dimensional
finite-temperature effective potential~\cite{Chen:2017qcz} in this work.

By varying the renormalisation scale of the 4-dimensional theory in the DR
matching relations, we can estimate the systematic uncertainty in our analysis arising from scale dependence.
By varying scale from $T$ to $4\pi e^{-\gamma}T \approx 7.05 T$ -- corresponding
to the average momentum of integration of the superheavy Matsubara modes
\cite{Farakos:1994kx} -- we find that we
have about 10 to 20 percent uncertainty in determination of $T_c$, $x_c$, and
the location of FOPT regions.

Furthermore, as in the original study of the SM \cite{Kajantie:1995dw}, also in
Ref.~\cite{Brauner:2016fla} for the xSM, the dimension-6 operators were
dropped from the effective theory. While it is difficult to estimate the effect
 of the dimension-6 operators comprehensively, we can include leading order
 contributions to the operator $c_6 (\phi_3^\dagger \phi_3)^3$ and analyze the
 effective potential with this operator in 3-d perturbation theory. A similar
 approach was recently used in Ref.~\cite{Gorda:2018hvi} (see Section 3.4.2 and
 Appendix C.7 therein for details).

Using the tools of Ref.~\cite{Brauner:2016fla}, we can estimate the matching
coefficient for the $(\phi_3^\dagger \phi_3)^3$ operator. In particular, utilizing the
 effective potential, Eqs. (3.27) and (3.29) in Ref.~\cite{Brauner:2016fla}, we
 obtain $V_{6,0} (\varphi^\dagger \varphi)^3$ with
\begin{multline}
V_{6,0} \simeq \Big(\frac{d}{16} g^6 + 40 \lambda^3 \Big)  I^{4b}_3 - 2 y^6_t I^{4f}_3 + \frac{1}{6} a^3_2 \tilde J^{4b}_3(\mu_\sigma) \\
= \frac{8 \zeta(3)}{3 (4\pi)^4 T^2}\Big( \frac{3}{64} g^6 - \frac{21}{2}y^6_t + 30 \lambda^3 \Big) + \frac{a^3_2}{6}\Big(\frac{1}{32\pi^2 \mu^2_\sigma} + J_3(\mu_\sigma) \Big),
\end{multline}
where $J_3(\mu_\sigma)\equiv 1/(2\pi^2 T^2) J_B'''(\mu^2_\sigma/T^2)$, where $J_B$, called $J_+$ in our analysis, is given e.g.~by Eq.~(B1) of Ref.~\cite{Niemi:2018asa}. The other relevant master integrals can be found in Appendix B of Ref.~\cite{Brauner:2016fla} and in Ref.~\cite{Niemi:2018asa}.
The first three terms above are the SM contributions and can be compared to Eqs.~(196) and (197) in \cite{Kajantie:1995dw}. We emphasize that whereas the sumintegral $I^{4b}_3$ contains only contributions from non-zero Matsubara modes, the sumintegral $\tilde J^{4b}_3(\mu_\sigma)$ contains the zero mode as well, and the mass is now assumed to be superheavy.
In the above expression, we have imposed a $Z_2$-symmetry for simplicity, i.e we set $b_1 = b_3 = a_1 = 0$. Without imposing a $Z_2$-symmetry, singlet contributions to the doublet 6-point function at one-loop become more complicated, as in addition to contributions from the effective potential, one needs to include all one-$\sigma$-reducible diagrams. However, in the non-$Z_2$-symmetric case, there are already contributions at tree-level
\begin{align}
\label{eq:dim-6-non-Z2}
\frac{1}{24} \Big( \frac{a^2_1 a_2}{\mu^4_\sigma} - \frac{a^3_1 b_3}{\mu^6_\sigma} \Big),
\end{align}
and we use these contributions to estimate the leading order effect. Therefore, our estimate of leading behavior reads
\begin{align} \label{eq:c6}
c_6 \simeq \frac{8 \zeta(3)}{3 (4\pi)^4}\Big( \frac{3}{64} g^6 - \frac{21}{2}y^6_t + 30 \lambda^3 \Big) + T^2 \frac{a^3_2}{6}\Big(\frac{1}{32\pi^2 \mu^2_\sigma} + J_3(\mu_\sigma) \Big) + \frac{T^2}{24} \Big( \frac{a^2_1 a_2}{\mu^4_\sigma} - \frac{a^3_1 b_3}{\mu^6_\sigma} \Big).
\end{align}
With this result, and using the Landau gauge effective potential of
Ref.~\cite{Gorda:2018hvi}, we observe that for the point in
Table~\ref{tab:sample-point} the critical temperature changes much less than one
percent when including this operator, while the scalar VEV changes about -20\%. In 3-d perturbation
theory, values for $\alpha$ and $\beta/H_*$ change about -40\% and -60\%,
 respectively. Overall, these estimates suggest that the uncertainty resulting from
 integrating out the singlet zero mode could be several tens of percent for points featuring relatively strong first-order transitions, which
 ultimately propagates to the nonperturbative analysis of the 3-d theory. 
 
For weaker first-order transitions, these effects are expected to be less significant. From the arguments of Ref.~\cite{Kajantie:1995dw}, the shift in the VEV scales approximately linearly with $c_6 / \lambda_{h,3}$. The SM dimension-6 operators cause a shift of around $1\%$ for values of $x$ corresponding to first-order transitions
 near the crossover region~\cite{Kajantie:1995dw}, and so we expect the change in the Higgs VEV to be $\mathcal{O}(c_6/c_{6, {\rm SM}} \times 1\%)$ for relatively weak transitions. On Figs.~\ref{fig:Z2}-\ref{fig:L221_mu3} we show where $c_6=10 c_{6, {\rm SM}}$, assuming $T\sim 140$ GeV. In the first-order transition regions for which this ratio is smaller than 10 or so, the corresponding shift in the VEV is expected be of order a few percent, and our nonperturbative analysis should yield results with accuracy comparable to that achieved in the original SM dimensional reduction and nonperturbative studies.  

\end{widetext}

\bibliography{singletprd}

\end{document}